\documentclass[12pt]{article}
\usepackage{feynmf}  
\renewcommand{\thefootnote}{\fnsymbol{footnote}}
\newcounter{line}

\def\ibid#1#2#3{{\it ibid.{} }{\bf #1} (#2) #3}
\def\ie{\hbox{\it i.e.}{}}
\def\eg{\hbox{\it e.g.}{}}
\def\etc{\hbox{\it etc}{}}
\def\nn{\hspace{2mm}}
\def\sss{\scriptscriptstyle}
\newcommand{\MeV}{\mbox{\rm MeV}}
\newcommand{\GeV}{\mbox{\rm GeV}}
\newcommand{\eV}{\mbox{\rm eV}}
\def\sleq{\raisebox{-.6ex}{${\textstyle\stackrel{<}{\sim}}$}}
\def\sgeq{\raisebox{-.6ex}{${\textstyle\stackrel{>}{\sim}}$}}
\def\Tr{{\rm Tr}{}}
\def\Tilde#1{\widetilde{#1}}

\def\Bar#1{\overline{#1}}

\def\ket#1{\left| #1\right\rangle}

\def\sVEV#1{\left\langle #1\right\rangle}

\def\abs#1{\left| #1\right|} 
\def\braket#1{\left\langle #1\right\rangle}
\def\cL{{\cal L}}
\def\AGUT{{}\;\;\raisebox{.9ex}{$\times$}\raisebox{-.5ex}%
{$\!\!\!\!\!\!\!\!\sss i=1,2,3$} \,(SMG_i \times U(1)_{\sss B-L,i})}%

\begin{document}
\begin{titlepage}
\begin{flushleft}
\vspace*{-1.5cm}
\vbox{\halign{#\hfil        
\cr
DESY 02-033    \cr
NBI-HE-02-02   \cr
hep-ph/0204027 \cr
April 2002  \cr
}}  
\end{flushleft}
\vspace*{-1cm}
\begin{center}
{\Large {\bf Baryogenesis via Lepton Number Violation \\ 
and Family Replicated Gauge Group}\\}

\vspace*{7mm}
{\ H. B. Nielsen}\footnote[1]{E-mail: hbech@mail.desy.de}
and {\ Y. Takanishi}\footnote[2]{E-mail: yasutaka@mail.desy.de}
            
\vspace*{.2cm}
{\it Deutsches Elektronen-Synchrotron DESY, \\
Notkestra{\ss}e 85,\\
D-22603 Hamburg, \\
Germany}\\
\vskip .15cm
{\it and}
\vskip .15cm
{\it The Niels Bohr Institute,\\
Blegdamsvej 17, \\
DK-2100 Copenhagen {\O}, \\
Denmark}\\
\vspace*{.3cm}
\end{center}
\begin{abstract}

We developed a previous model fitting all quark and lepton
-- including neutrino quantities in the region of the 
Large Mixing Angle-MSW solar solution -- order of magnitudewise by
only six adjustable parameters (Higgs vacuum expectation values) 
to also give an agreeing prediction for the amount of baryogenesis
produced in the early time of cosmology. We use Fukugita-Yanagida
scheme and take into account now also the effect of the remormalisation 
equation for the Dirac neutrino sector from Planck scale 
to the see-saw scale. The present version of our model with many 
approximately conserved (gauge) charges distinguishing various 
left- and right-Weyl particles has the largest matrix element of 
the mass matrix for the three flavour see-saw neutrinos being
the off-diagonal elements associated with the second- and 
third-proto-flavour and gives the ratio of baryon number density
to the entropy density to $2.59{+17.0\atop-2.25}\times 10^{-11}$ 
which agrees perfectly 
well as do also all the fermion masses and their mixing angles order
of magnitudewise.

\noindent\ Baryogenesis, Lepton number violation, Proton decay

\end{abstract}
\end{titlepage}
\newpage
\renewcommand{\thefootnote}{\arabic{footnote}}
\setcounter{footnote}{0}
\setcounter{page}{2}
\section{Introduction}\label{intro}
\indent\ Recently we have improved a rather specific model~\cite{FNT}
seeking to fit/explain quark and lepton masses and mixing angles
using as the reason for the large mass ratios and usually small
mixing angles, except two neutrino mixing angles, 
approximately conserved quantum numbers invented 
for that purpose into our model. The approximately conserved 
quantum numbers -- which are really assumed to be gauged, although 
that may not be so crucial -- are supposed to be broken by various 
Higgs fields. We could get a good fit of all the masses and 
mixing angles. However, this previous 
version of our model fails by providing too little 
baryogenesis to be produced in the early Big Bang. To calculate 
the baryogenesis we use the Fukugita-Yanagida scheme for 
producing a $B-L$ excess at the time when the temperature of 
the Universe passes the scale of the see-saw neutrino 
masses; in fact the see-saw mechanism for neutrinos masses is 
incorporated into our model in as far as we have three right-handed
neutrinos having Majorana masses acquired via a certain Higgs vacuum 
expectation value (VEV), $\sVEV{\phi_{\sss B-L}}$.

In such model(s) we can play around with the system of 
charges, but most importantly we have played with developing 
the charge assignments of the Higgs fields breaking the 
(gauge) group so that these charges are only 
\underline{approximately} conserved. It is the detailed 
quantum numbers of the speculated Higgses that we used to 
develop to fit more and more experimental informations. In the 
last improvement we got the model to predict Large Mixing 
Angle-MSW~\cite{MSW} (LMA-MSW) solution rather than Small 
Mixing Angle-MSW (SMA-MSW) solution by replacing the two 
of our Higgs fields which are able to cause transitions between
first and second generation quarks and leptons
by a new pair which actually turned out to have 
more elegant quantum numbers.

As we are now fitting all the masses and mixing angles, the 
constraints on our model are so tight that we hardly can 
change it anymore. However, we shall see below -- and that is the main 
point of the present article -- that we could still change 
the quantum numbers of the Higgs field delivering the over all scale
of the see-saw neutrino masses. 

The difficulty of getting enough baryons with our type of model 
hangs together 
with that even the largest element in the (Dirac) neutrino 
matrix is suppressed compared to unity by a factor $10^{-2}$.
Only our up-type quark and right-handed neutrino mass matrix have 
order of unity elements. Models with two Weinberg-Salam Higgs 
doublets can on the contrary have a $\tan\beta$ that is large 
and give suppressions to the down-type quark and charged lepton
masses so that the Yukawa couplings for these (lepton 
and down-type quark) particles could be as large as of order 
unity. That is to say two Higgs doublet models may in 
general tend to get stronger Yukawa couplings and thereby easier $CP$ 
violation since the latter comes only in one-loop accuracy.

It means that we are in need for possible enhancements for baryogenesis. The 
possibility for that which it actually turns out that we can obtain 
very elegantly in our model is to make the mass difference 
between some of the see-saw neutrinos smaller than in the 
previous model. In fact the $CP$ violating 
asymmetry parameter is essentially inversely proportional
to the mass difference between the see-saw neutrino producing 
the asymmetry in its decay and another see-saw neutrino. Due to 
the self-energy contribution almost degenerate neutrinos 
can even enhance the asymmetry by orders of 
magnitude compared to what it would be with the right-handed neutrino 
mass ratios of order unity. It would therefore be a progress in 
the direction of producing enough baryons to put two of our see-saw 
neutrinos almost degenerate in mass. It is very 
natural to obtain this situation by letting 
an off-diagonal element in the see-saw neutrino
mass matrix dominate. That may be done by adjusting the 
quantum numbers of that Higgs field $\phi_{\sss B-L}$ which 
gives the masses and the mass scale of the see-saw neutrinos 
so that it causes the transition from one flavour to another. 

This article is organised as follows: in the next section, 
we present general assumptions for the model building, and 
we review our model -- the family replicated gauge group model. In 
section $3$ we define our notation for the fermion 
mass terms in the Lagrangian and their mixing angles.
Then, in section $4$ the renormalisation group equations 
of all sectors are presented. The calculation 
is described in section $5$ and the results for the 
fermion masses and mixing angles are presented in section $6$. 
In section $7$ we discuss the Fukugita-Yanagida scheme of baryon 
number production, and in section $8$ the wash-out of 
the produced $B-L$ excess is approximated and 
our results of the calculation are given in section $9$. 
A discussion on the proton decay goes into section $10$.
Finally, section $11$ contains our conclusions.

\section{Model with many mass protecting charges}\label{muone}
\subsection{General feature}
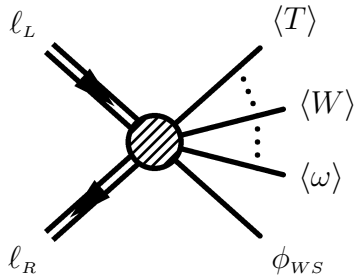
\begin{figure}[b!]
\vspace{5mm}
  \begin{center}
\def\F#1#2{\fmfi{dots}{.8[vloc(__z),vloc(__#1)] %
-- .8[vloc(__z),vloc(__#2)]}}
\parbox{35mm}{
\unitlength=1mm
\begin{fmffile}{higgsfig}
\begin{fmfgraph*}(35,25)
\fmfpen{thick}
\fmfleftn{i}{4}
\fmfrightn{j}{4}
\fmf{phantom}{i2,z,i3}
\fmf{dbl_plain_arrow}{i4,z,i1}
\fmf{plain}{j4,z,j3}
\fmf{plain}{j2,z,j1}                                                           
\fmffreeze
\fmfv{dec.shape=circle,dec.fill=shaded, dec.size=.2w}{z}
\fmfdraw
\F{j2}{j3}
\F{j3}{j4}
\fmflabel{$\ell_{\sss L}$}{i4}
\fmflabel{$\ell_{\sss R}$}{i1}
\fmflabel{$\phi_{\sss WS}$}{j1}
\fmflabel{$\sVEV{\omega}$}{j2}
\fmflabel{$\sVEV{T}$}{j4}
\fmflabel{$\sVEV{W}$}{j3}
\end{fmfgraph*}\end{fmffile}}
\vspace{12mm}
\caption{Diagram giving the effective Standard Model Dirac Yukawa coupling. 
Here Weinberg-Salam Higgs field is denoted as $\phi_{\sss WS}$.}
\label{diagram}
\end{center}
\end{figure}

\indent\ The main point of the model we use here is that the ``small
hierarchies'' of the quark and lepton masses are due to approximately
conserved quantum numbers~\cite{FN} preventing the masses from being different 
from zero, $\ie$, that the Yukawa couplings which are in Standard Model 
mysteriously small should really be understood as being so due to that 
they need breaking of some charge conservations. In fact we 
consider these Yukawa couplings just some effective 
couplings representing at the fundamental level more complicated 
vertex diagrams involving attachments to the VEVs of Higgs 
fields breaking the charge conservations in question
(see Fig.~\ref{diagram}).  

We summarise here our philosophy to build a model that can predict,
rather say fits by it having the all fermion quantities 
including not only the neutrino
oscillations but also baryogenesis in following ingredients:
\begin{itemize}
\item All fundamental coupling constants and masses 
are of order unity.
\begin{itemize}
\item[] All gauge and Yukawa couplings are of order one at the 
Planck scale, in fact, we take the fundamental scale as the Planck one.
{}From this point of view we treat them as random numbers,
and partly to be able to ``explain'' $CP$ violation we assume
that all Yukawa couplings are also complex random numbers. Also 
the masses of particles should be of order Planck scale
unless they are mass protected, so as to have masses determined by 
the VEVs of Higgs fields say.
\end{itemize}
\item Only VEVs are allowed far from order one.
\begin{itemize}
\item[] Thus also the scalar field masses corresponding to the 
VEVs are allowed to be very small, since otherwise it would be 
inconsistent to have $M_H^2=\lambda\,\sVEV{\phi_{\sss WS}}$ 
unless $\lambda$ has huge value.
\end{itemize}
\begin{description}
\item This assumption could gain supports from 
\item[{\it (a)}] the well-known weak scale being extremely 
small compare to the Planck scale.
\item[{\it (b)}] dynamical symmetry $\eg$ Nambu-Jona-Lasinio~\cite{NJL}
can lead to equation for the logarithm of the VEV scale.
\item[{\it (c)}] in superconductivity VEVs tend very small compared to the
atomic dimensionality argument. 
\end{description}
\item There are {\it a priori} several conserved charges and the 
various Weyl components of quarks and leptons have usually different 
charges, $\ie$, quantum numbers.
\begin{itemize}
\item[] Since we let Weyl components of different families have in 
general different quantum numbers of these kinds, they are to be 
considered horizontal, but they should better not
bring together into multiplets different families because that 
would make the remarkably large inter-family mass ratios difficult 
to incorporate (unless the group is strongly broken).  
\end{itemize}
\item Incorporate the see-saw mechanism~\cite{seesaw}.
\begin{itemize}
\item[] A new scale must be introduced into the model. In our case we assume
a new scale about $10^{10}~\GeV$. This scale becomes crudely the masses of
right-handed (Majorana) neutrinos, and therefore $B-L$ quantum charge 
being broken at that energy scale.
\end{itemize}
\end{itemize}
Note that we do not assume the existence of Supersymmetry
or Grand Unification. However, our prediction came out
that $SU(5)$ mass relations are {\em order of magnitudewise}
valid due to the diagonal
elements of different mass matrices need same quantum numbers
being broken in our specific model.

\subsection{Our specific model of many conserved charges}\label{ourmodel}
\indent\ We have already investigated a model~\cite{FNT,NT} which can 
predict all quark and charged lepton quantities including 
baryogenesis\footnote{The model~\cite{NT} provided only the SMA-MSW 
solution. A main point of present article is to calculate baryogenesis
with LMA-MSW solution predicting mass matrices including running effects 
of all sectors (see sections~\ref{RGE}, $7$-$9$).}. This model 
is of the type described in the foregoing 
subsection with the further good restriction that the quantum numbers 
for the Weyl particles are so chosen that they obey the 
anomaly cancelation conditions required for them to 
be imagined gauged charges. It is namely easily seen that our
gauge group has no anomalies -- neither gauge nor mixed ones -- because 
it has the Standard Model group plus a $B-L$ separately for each 
family. In fact our model has the specific group 
\begin{equation}
  \label{eq:agut}
  \AGUT\nn,
\end{equation}
where $SMG_i\equiv SU(3)_i\times SU(2)_i\times U(1)_i$ denotes 
the Standard Model gauge group for each generation ($i=1,2,3$); 
$\times$ means the Cartesian product. However, {\em it is unlikely 
to be important to have this special group, provided we have sufficiently many 
conserved charges separating in quantum number the various Weyl fields}. 
Actually the important thing is that one by means
of the quantum numbers can construct some Higgs fields as we propose them 
for the breaking of the group to the Standard Model with much the same 
powers of fields being needed in the various mass matrix elements.
The latter may in fact not be difficult to obtain with other groups, at 
least the non-abelian part 
of our proposed group is totally irrelevant for the fitting. 

We should emphasize that due to the non-zero neutrino masses using 
the see-saw picture, it is necessary to introduce a right-handed neutrino, 
or preferably several -- we have three in our model -- having Majorana masses.
In our model we associate the see-saw mass scale with a gauged $B-L$ 
charge. We let the latter come from the diagonal subgroup
of the Cartan product of a $U(1)_{\sss B-L,i}$ ($i=1,2,3$) for each family.

Note that this family replicated gauge group, eq.~(\ref{eq:agut}), 
is the maximal gauge group under the following assumptions:
\begin{list}{(\arabic{line})}{\usecounter{line}}
\item Considering only part of the gauge group acting
non-trivially on the known 45 Weyl fermions of the Standard Model 
and the additional three heavy see-saw (right-handed) neutrinos, $\ie$,
our gauge group is a subgroup of $U(48)$.
\item Avoid gauge transformations transforming a 
Weyl state from one irreducible representation of the Standard Model 
group into another irreducible representation, $\ie$, there is no 
place for the quarks and leptons occupying the same irreducible 
representation under our gauge group.
\item  No gauge nor mixed anomalies due to the renormalisability 
requirement.
\end{list}

\subsection{Description of quantum numbers of mass protected particles 
in model}\label{ourmodel2}
\indent\ In addition to the practically any non-mass-protected 
particles, which 
we have already assumed to be present with Planck scale masses, 
we have as is easily seen some mass protected particles. The 
Weyl particles without chiral partners in our model are 
the well-known Weyl components of the quarks and leptons in the 
Standard Model plus three see-saw (right-handed) heavy neutrinos, 
which are all easily seen to be mass protected
by our gauge group~eq.(\ref{eq:agut}). These $48$ Weyl particles 
have lower mass than the Planck scale. They fall into three 
families each consisting of the $16$ Weyl particles of a 
usual Standard Model generation plus one see-saw particle. In 
this way we can label these particle as proto-left-handed or
proto-right-handed $u$-quark, $d$-quark, electron $\etc$. To get the 
quantum numbers under our model gauge group for a given 
proto-irreducible representation, we proceed in the following way:
We note the generation number $i~(i=1,2,3)$ of the particle for which we want 
quantum numbers and we look up, in the Standard Model, what are the 
quantum numbers of the irreducible representation in question 
and what is the $B-L$ quantum number. Then we take trivial quantum numbers
for the two generations $j$ different from $i$ and put the index $i$ 
to the gauge group's names, so we put the representations 
of the groups $SU(3)_i$, $SU(2)_i$ and $U(1)_i$ equal to the 
just asked quantum numbers. 

{}For instance, if we want to find 
the quantum numbers of the proto-left-handed bottom quark, we note 
that the quantum numbers of the left-handed bottom quark in the 
Standard Model are weak hypercharge $y/2=1/6$, 
doublet under $SU(2)$ and triplet under $SU(3)$, 
while $B-L$ is equal to the baryon number $=1/3$. Moreover, 
ignoring mixing angles, the 
generation is denoted as number $i=3$.
The latter fact means that all the quantum numbers for $SMG_j$ 
$j=1,2$ are trivial. Also the baryon number minus lepton number for the
proto-generation number one and two are zero: only the quantum numbers 
associated with proto-generation three are non-trivial. Thus, in our model, 
the quantum numbers of the proto-left-handed bottom quark are 
$y_3/2 = 1/6$, doublet under $SU(2)_3$, triplet under 
$SU(3)_3$ and $(B-L)_3=1/3$. For each proto-generation the following charge 
quantisation rule applies
\begin{equation}
  \label{eq:mod}
  \frac{t_i}{3}+\frac{d_i}{2} + \frac{y_i}{2} = 0~~{\rm (mod~1)}\nn,
\end{equation}
where $t_i$ and $d_i$ are the triality and duality for 
the $i$'th proto-generation gauge groups $SU(3)_i$ and $SU(2)_i$ 
respectively.

Combining eq.~(\ref{eq:mod}) with the principle of taking the smallest 
possible representation of the groups $SU(3)_i$ and $SU(2)_i$, it 
is sufficient to specify the six Abelian quantum numbers $y_i/2$ 
and $(B-L)_i$ in order to completely specify the gauge quantum 
numbers of the fields, $\ie$ of the Higgs fields and fermion fields.
Using this rule we easily specify the fermion representations as in 
Table~\ref{Table1}. However, as already mentioned in section $2.2$ 
these non-abelian $SU(3)_i$ and $SU(2)_i$ are really not important 
for our fit since they just follow the weak hypercharges 
like ``slaves''.

Note that fermion quantum numbers for each proto-generation gauge group 
$SMG_i \times U(1)_{B-L,i}$ are obtainable as considering it  a subgroup of 
$SO(10)$, $\ie$ our gauge group eq.~(\ref{eq:agut}) is 
really a subgroup of $SO(10)^3$. However, we avoid the not well-agreeing
accurate predictions of simplest grand unification by {\em not having neither 
$SU(5)$'s nor $SO(10)$'s:} the mass spectra, and also proton decay.

\begin{table}[!ht]
\vspace{3mm}
\label{Table1}
\begin{center}
\begin{tabular}{|c||c|c|c|c|c|c|} \hline
& $SMG_1$& $SMG_2$ & $SMG_3$ & $U_{\sss B-L,1}$ & $U_{\sss B-L,2}$ & 
$U_{\sss B-L,3}$ \\ \hline\hline
$u_L,d_L$ &  $\frac{1}{6}$ & $0$ & $0$ & $\frac{1}{3}$ & $0$ & $0$ \\
$u_R$ &  $\frac{2}{3}$ & $0$ & $0$ & $\frac{1}{3}$ & $0$ & $0$ \\
$d_R$ & $-\frac{1}{3}$ & $0$ & $0$ & $\frac{1}{3}$ & $0$ & $0$ \\
$e_L, \nu_{e_{\sss L}}$ & $-\frac{1}{2}$ & $0$ & $0$ & $-1$ & $0$ & $0$ \\
$e_R$ & $-1$ & $0$ & $0$ & $-1$ & $0$ & $0$ \\
$\nu_{e_{\sss R}}$ &  $0$ & $0$ & $0$ & $-1$ & $0$ & $0$ \\ \hline
$c_L,s_L$ & $0$ & $\frac{1}{6}$ & $0$ & $0$ & $\frac{1}{3}$ & $0$ \\
$c_R$ &  $0$ & $\frac{2}{3}$ & $0$ & $0$ & $\frac{1}{3}$ & $0$ \\
$s_R$ & $0$ & $-\frac{1}{3}$ & $0$ & $0$ & $\frac{1}{3}$ & $0$\\
$\mu_L, \nu_{\mu_{\sss L}}$ & $0$ & $-\frac{1}{2}$ & $0$ & $0$ & $-1$ & 
$0$\\ $\mu_R$ & $0$ & $-1$ & $0$ & $0$  & $-1$ & $0$ \\
$\nu_{\mu_{\sss R}}$ &  $0$ & $0$ & $0$ & $0$ & $-1$ & $0$ \\ \hline
$t_L,b_L$ & $0$ & $0$ & $\frac{1}{6}$ & $0$ & $0$ & $\frac{1}{3}$ \\
$t_R$ &  $0$ & $0$ & $\frac{2}{3}$ & $0$ & $0$ & $\frac{1}{3}$ \\
$b_R$ & $0$ & $0$ & $-\frac{1}{3}$ & $0$ & $0$ & $\frac{1}{3}$\\
$\tau_L, \nu_{\tau_{\sss L}}$ & $0$ & $0$ & $-\frac{1}{2}$ & $0$ & $0$ & 
$-1$\\ $\tau_R$ & $0$ & $0$ & $-1$ & $0$ & $0$ & $-1$\\
$\nu_{\tau_{\sss R}}$ &  $0$ & $0$ & $0$ & $0$ & $0$ & $-1$ \\ 
\hline \hline
$\omega$ & $\frac{1}{6}$ & $-\frac{1}{6}$ & $0$ & $0$ & $0$ & $0$\\
$\rho$ & $0$ & $0$ & $0$ & $-\frac{1}{3}$ & $\frac{1}{3}$ & $0$\\
$W$ & $0$ & $-\frac{1}{2}$ & $\frac{1}{2}$ & $0$ & $-\frac{1}{3}$ 
& $\frac{1}{3}$ \\
$T$ & $0$ & $-\frac{1}{6}$ & $\frac{1}{6}$ & $0$ & $0$ & $0$\\
$\chi$ & $0$ & $0$ & $0$ & $0$ & $-1$ & $1$ \\ 
$\phi_{\sss WS}$ & $0$ & $\frac{2}{3}$ & $-\frac{1}{6}$ & $0$ & 
$\frac{1}{3}$ &  $-\frac{1}{3}$ \\
$\phi_{\sss B-L}$ & $0$ & $0$ & $0$ & $0$ & $1$ & $1$ \\
\hline
\end{tabular}
\end{center}
\caption{All $U(1)$ quantum charges for the proto-fermions 
and Higgs fields in the model.}
\end{table}
\subsection{Higgses breaking the family replicated gauge group to the 
Standard Model}
\indent
The gauge group $\AGUT$ is at first spontaneously broken down at one 
or two orders of magnitude below the Planck scale, by 5 different 
Higgs fields, to the gauge group $SMG\times U(1)_{B-L}$ which is 
the diagonal subgroup of the original one:
\begin{equation}
  \label{eq:subagut}
\AGUT \rightarrow \left(~\AGUT~\right)_{\rm diagonal}\equiv SMG\times U(1)_{B-L}\nn.
\end{equation}
This diagonal subgroup is further broken down by yet two more 
Higgs fields --- the Weinberg-Salam Higgs field $\phi_{WS}$ and 
another Higgs field $\phi_{B-L}$ --- to 
$SU(3)\times U(1)_{em}$. The VEV of 
the $\phi_{B-L}$ Higgs field is taken to be about $10^{10}~\GeV$ 
and is designed to break the gauged $B-L$ quantum number. In 
other words the VEV, $\sVEV{\phi_{B-L}}$, gives the see-saw scale
(All $U(1)$ quantum charged for the proto-fermions and 
Higgs fields are presented in Table~\ref{Table1}.).

We should emphanise here that the quantum numbers of the
$\phi_{B-L}$ has been changed relative to our previous publication due to 
the baryogenesis point 
of view. The point is that baryogenesis calculated the way described below
with the ``old'' quantum charge of $\phi_{B-L}$:
\begin{equation}
\vec{Q}_{\phi_{B-L}}~\Big|_{\rm~old} = (0,0,0,0,0,2)
\end{equation}
would produce insufficient amount of baryon minus lepton number, whereas
we shall see that quantum numbers making the off-diagonal 
elements $(2,3)$ and $(3,2)$ of the right-handed neutrino mass matrix 
be the dominant ones as in Table~\ref{Table1}:
\begin{equation}
\vec{Q}_{\phi_{B-L}} = (0,0,0,0,1,1) \nn.
\end{equation}

\section{{}Fermion masses and their mixing angles}\label{definitions}
\indent
In this section we define our notations for the mass terms
of the charged fermion sectors and the neutrino 
sectors. These define also the mixing angle unitary matrices
for quark sector and neutrino sector, respectively.

\subsection{Charged fermion masses and their mixing angles}
\indent

The full Lagrangian of our model is that of a gauge 
theory with scalars and Weyl particles of which many 
are the ones at the Planck scale with which we do not go 
into details. We shall therefore leave it to the readers 
imagination and only present the notation for the Weinberg-Salam Yukawa 
part effective Lagrangian of the Standard Model form
\begin{equation}
-{\cal L}_{{\rm charged-fermion-mass}} = 
\overline{Q_{\sss L}}{}Y_{\sss U} \tilde\Phi_{\sss WS}{}U_{\sss R} +        
\overline{Q_{\sss L}}{}Y_D\Phi_{\sss WS}{}D_{\sss R} +
        \overline{L_{\sss L}}{}Y_{\sss E}\Phi_{\sss WS}{}E_{\sss R} + h.c.
\label{L_Higgs}
\end{equation}
Here $\Phi_{WS}$ is the Weinberg-Salam Higgs field,
$Q_L$ denotes the three $SU(2)$ doublets of
left-handed quarks, $U_R$ denotes the three singlets of right-handed 
up-type quarks and $Y_U$ is the three-by-three Yukawa coupling 
matrix for the up-type quarks. Similarly $Y_D$ and $Y_E$ are 
the Yukawa coupling matrices for the down-type quarks and charged 
leptons respectively. The $SU(2)$ doublets $\Phi_{WS}$ and $Q_L$ 
can be represented
as $2$ component column vectors and we then define:
\begin{equation}
\tilde\Phi_{\sss WS} = \left ( \begin{array}{cc} 0 & 1 \\ 1 & 0 
\end{array}\right ) \Phi_{\sss WS}^{\dagger} 
\end{equation}
and
\begin{equation}
\overline{Q_{\sss L}} =
        \overline{ \left ( \begin{array}{c} U_{\sss L} \\
D_{\sss L} \end{array} \right ) }
        = ( \overline{U_{\sss L}} \; \overline{D_{\sss L}} ) \nn,
\end{equation}
where $\overline{U_{\sss L}}$ are the $CP$ conjugates of the three 
left-handed up-type quarks. 

The effective Yukawa couplings $Y_E$, for example, are obtained 
from identifying the lowest order vertex in eq.~(\ref{L_Higgs}) 
with a vertex Fig.~\ref{diagram}.

After electroweak symmetry breaking 
the Weinberg-Salam Higgs field gets a VEV and we obtain the following 
mass terms in the Lagrangian:
\begin{equation}
-{\cal L}_{{\rm charged-fermion-mass}} = \overline{U_{\sss L}} 
\,M_{\sss U}\, 
U_{\sss R} + \overline{D_{\sss L}} \,M_{\sss D}\, D_{\sss R} +
        \overline{E_{\sss L}} \,M_{\sss E}\, E_{\sss R} + h.c.
\label{L_Mass}
\end{equation}
where the mass matrices are related to the Yukawa coupling matrices
and Weinberg-Salam Higgs VEV by:
\begin{equation}
M = Y\; \frac{\sVEV{\phi_{\sss WS}}}{\sqrt{2}} \nn.
\label{mass-scale}
\end{equation}
We have chosen the normalisation from the Fermi coupling constant:
\begin{equation}
\sVEV{\phi_{\sss WS}} = 246\;\GeV\nn.
\label{WS-vev}
\end{equation}
In order to obtain the masses from the mass matrices, $M_{\sss U}$, 
$M_{\sss D}$ and $M_{\sss E}$, we must diagonalise them to find 
their eigenvalues. In particular we can find
unitary matrices, $V_{\sss U}$ for the up-type quarks, $V_{\sss D}$ 
for the down-type quarks and $V_{\sss E}$ for the charged leptons:
\begin{eqnarray}
V_{\sss U}^{\dagger}\,M_{\sss U}\,M_{\sss U}^{\dagger}\,V_{\sss U} & = &
\mbox{diag}\left( m_u^2, m_c^2, m_t^2 \right) \nn,\\
V_{\sss D}^{\dagger}\,M_{\sss D}\,M_{\sss D}^{\dagger}\,V_{\sss D} & = &
\mbox{diag}\left( m_d^2, m_s^2, m_b^2 \right) \nn,\\
\label{uni}
V_{\sss E}^{\dagger}\,M_{\sss E}\,M_{\sss E}^{\dagger}\,V_{\sss E} & = &
\mbox{diag}\left( m_e^2, m_\mu^2, m_\tau^2 \right) \nn.
\end{eqnarray}
The quark mixing matrix is then defined with these unitary matrices 
as~\cite{CKM}: 
\begin{equation}
V_{{\rm\sss CKM}} = V_{\sss U}^{\dagger}\, V_{\sss D} \nn.
\end{equation}
This $V_{{\rm\sss CKM}}$ unitary matrix is parameterised as follows:
\begin{eqnarray}
V_{\rm\sss CKM}&=& \left( \begin{array}{ccc}
  c_{12} c_{13}       & s_{12} c_{13}  & s_{13} e^{-i\delta_{13}} \\
- s_{12} c_{23} - c_{12} s_{23} s_{13} e^{i\delta_{13}}
& c_{12} c_{23} - s_{12} s_{23} s_{13} e^{i\delta_{13}}
& s_{23} c_{13} \\
    s_{12} s_{23} - c_{12} c_{23} s_{13} e^{i\delta_{13}}
& - c_{12} s_{23} - s_{12} c_{23} s_{13} e^{i\delta_{13}}
& c_{23} c_{13} \\
\end{array} \right)\nonumber\\
&=& \left( \begin{array}{ccc}
  V_{ud} & V_{us} & V_{ub}\\
  V_{cd} & V_{cs} & V_{cb}\\
  V_{td} & V_{ts} & V_{tb}\\
\end{array} \right) \nn,
\label{eq:CKM}
\end{eqnarray}
where $c_{ij} \equiv \cos\theta_{ij}$ and
$s_{ij} \equiv \sin\theta_{ij}$ for the generation labels 
$i,j=1,2,3$ and $\delta_{13}$ is a $CP$ violating
phase in the quark sector. 

\subsection{Neutrino masses and their mixing angles}\label{sec:seesaw}
\indent
The left-handed neutrinos are massless in Standard Model because
of non-existence of right-handed neutrinos
and the Weinberg-Salam Higgs field having only weakhypercharge
$y/2=1/2$, and not $y/2=1$ as is required to give Majorana neutrino
a mass. Thus, in order to let left-handed neutrinos be massive
particles as is very strongly indicated by experiments, we 
assume the existence of the three very heavy 
right-handed neutrinos, which are mass-protected so as finally 
get masses from the ``new'' Higgs field, $\phi_{B-L}$, at an energy scale of 
about $10^{10}~\GeV$ in our present model. 

We use the gauged $B-L$ charge to mass-protect
the right-handed neutrinos; in fact we use the total  -- 
diagonal -- one in fact we break $U(1)_{B-L, 1}\times U(1)_{B-L, 2}
\times U(1)_{B-L, 3}$ $\supset U(1)_{B-L}$ at a much higher 
energy scale, say about $10^{18}~\GeV$. 

The assumption of the existence of three right-handed Majorana 
neutrinos at a high scale gives rise to the addition of 
Majorana mass terms to the Lagrangian:
\begin{eqnarray}
  \label{lagrangian}
-\cL_{\sss\rm neutrino-mass} &\!=\!&
\bar{\nu}_L \,M^D_\nu \,\nu_R
+  \frac{1}{2}(\Bar{\nu_L})^{\sss c}\,M_{L}\, \nu_L
+  \frac{1}{2}(\Bar{\nu_R})^{\sss c}\,M_{R}\, \nu_R + h.c. \nonumber\\
&\!=\!&
\frac{1}{2} (\Bar{n_L})^{\sss c} \,M\, n_L + h.c.
\end{eqnarray}
where
\begin{equation}
n_L \!\equiv\! \left( \begin{array}{c}
    \nu_L \\
    (\nu_L)^{\sss c}
    \end{array} \right) \nn,\nn
M \!\equiv\! \left( \begin{array}{cc}
    M_L & M^D_\nu\\
    M^D_\nu & M_R
    \end{array} \right) \nn.
\end{equation}
\noindent
Here $M_\nu^D$ is the left-right transition mass term -- Dirac neutrino
mass term -- and  $M_L$ and $M_R$ are the isosinglet Majorana mass
terms of left-handed and right-handed neutrinos, respectively.  

Due to mass-protection by the Standard Model gauge symmetry, the 
left-handed Majorana mass terms, $M_L$, are negligible in our model 
with a fundamental scale set by the Planck mass. Then, 
naturally, the effective light (left-left transition) 
neutrino mass matrix can be obtained via the see-saw 
mechanism~\cite{seesaw}:
\begin{equation}
  \label{eq:meff}
  M_{\rm eff} \! \approx \! M^D_\nu\,M_R^{-1}\,(M^D_\nu)^T\nn.
\end{equation}

In the framework of the three active neutrinos model, the flavour
eigenstates $\nu_{\alpha}~(\alpha=e,\nu,\tau)$ are related to
the mass eigenstates $\nu_{i}~(i=1,2,3)$ in the vacuum by a 
unitary matrix $V_{\rm\sss MNS}$,
\begin{equation}
\label{eigen}
  \ket{\nu_{\alpha}} = \sum_i (V_{\rm\sss MNS})_{\alpha i} \ket{\nu_i}\nn.
\end{equation}
Here $V_{\rm\sss MNS}$ is the three-by-three 
Maki-Nakagawa-Sakata (MNS) mixing
matrix~\cite{MNS} which is parameterised by
\begin{eqnarray}
V_{\rm\sss MNS}&=& \left( \begin{array}{ccc}
  \tilde{c}_{12} \tilde{c}_{13} 
& \tilde{s}_{12} \tilde{c}_{13}  
& \tilde{s}_{13} e^{-i\tilde{\delta}_{13}} \\
- \tilde{s}_{12} \tilde{c}_{23} - \tilde{c}_{12} \tilde{s}_{23} \tilde{s}_{13} e^{i\tilde{\delta}_{13}}
& \tilde{c}_{12} \tilde{c}_{23} - \tilde{s}_{12} \tilde{s}_{23} \tilde{s}_{13} e^{i\tilde{\delta}_{13}}
& \tilde{s}_{23} \tilde{c}_{13} \\
    \tilde{s}_{12} \tilde{s}_{23} - \tilde{c}_{12} \tilde{c}_{23} \tilde{s}_{13} e^{i\tilde{\delta}_{13}}
& - \tilde{c}_{12} \tilde{s}_{23} - \tilde{s}_{12} \tilde{c}_{23} \tilde{s}_{13} e^{i\tilde{\delta}_{13}}
& \tilde{c}_{23} \tilde{c}_{13} \\
\end{array} \right)\nonumber\\
&&\nn\times\left( \begin{array}{ccc}
 e^{i\varphi} & 0 & 0 \\
 0 &  e^{i\psi} & 0\\
 0 &  0          & 1 \\
\end{array} \right) \nn,
\label{eq:MNS}
\end{eqnarray}
where $\tilde{c}_{ij} \equiv \cos\theta_{ij}$ and
$\tilde{s}_{ij} \equiv \sin\theta_{ij}$ are the neutrino mixing parameters
and $\tilde{\delta}_{13}$ 
is a neutrino $CP$ violating
phase. Note that, due to the existence of Majorana neutrinos, we
have two additional $CP$ violating Majorana phases $\varphi$ and 
$\psi$, which are also included in the MNS unitary mixing matrix.

In order to get predictions for the neutrino masses from the 
effective mass matrix, $M_{\rm eff}$, we have to diagonalise this 
matrix using a unitary matrix, $V_{\rm eff}$, to find the mass 
eigenvalues:
\begin{equation}
  \label{eq:mixingmatrix}
V_{\rm eff} M_{\rm eff} M_{\rm eff}^\dagger V_{\rm eff}^\dagger 
= {\rm diag}(m^2_1, m^2_2, m^2_3)\nn.
\end{equation}      
With the charged lepton unitary matrix $V_{\sss E}$, eq.~(\ref{uni}), 
we can then find the neutrino mixing matrix: 
\begin{equation}
  V_{\rm\sss MNS} = V_{\sss E}\,V_{\rm eff}^\dagger \nn.
\end{equation}%
Obviously, we should compare these theoretical predictions with
experimentally measured quantities, therefore we define:
\begin{eqnarray}
\label{eq:thtoex}
\Delta m^2_{\odot}&\equiv&m^2_2 - m^2_1\nn,\\
\Delta m^2_{\rm atm}&\equiv& m^2_3 - m^2_2 \nn,\\
\tan^2\theta_{\odot} &\equiv& \tan^2\theta_{12} \nn,\\
\tan^2\theta_{\rm atm}&\equiv& \tan^2\theta_{23}\nn, \\      
\tan^2\theta_{\rm chooz}&\equiv& \tan^2\theta_{13} \nn.
\end{eqnarray}

Note that since we use the philosophy of order of 
magnitudewise predictions (see section~\ref{sec:FN}) with 
complex order one coupling constants, 
our model is capable of making predictions for these
three phases, the $CP$ violating phase $\tilde{\delta}_{13}$ 
and the two Majorana phases; put
simply, we assume that all these phases are of order $\pi/2$, 
$\ie$ essentially maximal $CP$ violations.

\section{Renormalisation group equations}\label{RGE}
\indent
{}From the Planck scale down to the see-saw scale or rather from 
where our gauge group is broken down to $SMG\times U(1)_{B-L}$ we use
the one-loop renormalisation group running of the Yukawa coupling constant 
matrices and the gauge couplings~\cite{pierre} including the running
of Dirac neutrino Yukawa coupling:
\begin{eqnarray}
\label{eq:recha}
16 \pi^2 {d g_{1}\over d  t} &\!=\!& \frac{41}{10} \, g_1^3 \nn,\\
16 \pi^2 {d g_{2}\over d  t} &\!=\!& - \frac{19}{16} \, g_2^3 \nn, \\
16 \pi^2 {d g_{3}\over d  t} &\!=\!& - 7 \, g_3^3  \nn,\\
16 \pi^2 {d Y_{\sss U}\over d  t} &\!=\!& \frac{3}{2}\, 
\left( Y_{\sss U} (Y_{\sss U})^\dagger
-  Y_{\sss D} (Y_{\sss D})^\dagger\right)\, Y_{\sss U} 
+ \left\{\, Y_{\sss S} - \left(\frac{17}{20} g_1^2 
+ \frac{9}{4} g_2^2 + 8 g_3^2 \right) \right\}\, Y_{\sss U}\nn,\\
16 \pi^2 {d Y_{\sss D}\over d  t} &\!=\!& \frac{3}{2}\, 
\left( Y_{\sss D} (Y_{\sss D})^\dagger
-  Y_{\sss U} (Y_{\sss U})^\dagger\right)\,Y_{\sss D} 
+ \left\{\, Y_{\sss S} - \left(\frac{1}{4} g_1^2 
+ \frac{9}{4} g_2^2 + 8 g_3^2 \right) \right\}\, Y_{\sss D}\nn,\\
16 \pi^2 {d Y_{\sss E}\over d  t} &\!=\!& \frac{3}{2}\, 
\left( Y_{\sss E} (Y_{\sss E})^\dagger
-  Y_{\sss \nu} (Y_{\sss \nu})^\dagger\right)\,Y_{\sss E} 
+ \left\{\, Y_{\sss S} - \left(\frac{9}{4} g_1^2 
+ \frac{9}{4} g_2^2 \right) \right\}\, Y_{\sss E} \nn,\\
\label{Diracyukawa}
16 \pi^2 {d Y_{\sss \nu}\over d  t} &\!=\!& \frac{3}{2}\, 
\left( Y_{\sss \nu} (Y_{\sss \nu})^\dagger
-  Y_{\sss E} (Y_{\sss E})^\dagger\right)\,Y_{\sss \nu} 
+ \left\{\, Y_{\sss S} - \left(\frac{9}{20} g_1^2 
+ \frac{9}{4} g_2^2 \right) \right\}\, Y_{\sss \nu} \nn,\\
 \label{YScon} Y_{\sss S} &\!=\!& {\Tr}(\, 3\, Y_{\sss U}^\dagger\, Y_{\sss U} 
+  3\, Y_{\sss D}^\dagger \,Y_{\sss D} +  Y_{\sss E}^\dagger\, 
Y_{\sss E} +  Y_{\sss \nu}^\dagger\, Y_{\sss \nu}\,) \nn,
\end{eqnarray}
where $t=\ln\mu$ and $\mu$ is the renormalisation point.

However, below the see-saw scale the right-handed neutrino
are no more relevant and the Dirac neutrino terms in the 
renormalisation group equations should be removed, and also 
the Dirac neutrino Yukawa couplings themselves are not accessible
anymore. That means that, from the see-saw scale down to the experimental
scale ($1~\GeV$), the only leptonic Yukawa $\beta$-functions
should be changed as follows:
\begin{equation}
16 \pi^2 {d Y_{\sss E}\over d  t} =\frac{3}{2}\, 
\left( Y_{\sss E} (Y_{\sss E})^\dagger \right)\,Y_{\sss E} 
+ \left\{\, Y_{\sss S} - \left(\frac{9}{4} g_1^2 
+ \frac{9}{4} g_2^2 \right) \right\}\, Y_{\sss E} \nn.
\end{equation}

Note that the quantity, $Y_{\sss S}$, must be also changed
below the see-saw scale:
\begin{equation}
\label{eq:Y_S}
Y_{\sss S}={\Tr}(\, 3\, Y_{\sss U}^\dagger\, Y_{\sss U} 
+  3\, Y_{\sss D}^\dagger \,Y_{\sss D} +  Y_{\sss E}^\dagger\, 
Y_{\sss E}\,)  \nn.
\end{equation}

{}Further, we should evolve 
the effective neutrino mass matrix considered as a 
whole running as an irrelevant or nonrenormalisable $5$ 
dimensional term~\cite{5run} from the see-saw 
scale, set by $\sVEV{\phi_{B-L}}$ to our model, to $1~\GeV$:
\begin{equation}
\label{eq:remeff}
16 \pi^2 {d M_{\rm eff} \over d  t}
= ( - 3 g_2^2 + 2 \lambda + 2 Y_{\sss S} ) \,M_{\rm eff}
- {3\over 2} \left( M_{\rm eff}\, (Y_{\sss E} Y_{\sss E}^\dagger) 
+ (Y_{\sss E} Y_{\sss E}^\dagger)^T \,M_{\rm eff}\right) \nn,
\end{equation}
where $Y_{\sss S}$ defined in eq.~(\ref{eq:Y_S}) and in this energy range
the Higgs self-coupling constant running equation is
\begin{equation}
\label{eq:rehiggs}
16 \pi^2 {d \lambda\over d  t}
= 12 \lambda^2 - \left( \frac{9}{5} g_1^2 + 9 g_2^2 \right) \,\lambda
+ \frac{9}{4} \left( \frac{3}{25} g_1^4 
+ \frac{2}{5} g_1^2 g_2^2 + g_2^4 \right) + 4 Y_{\sss S} \lambda 
- 4 H_{\sss S}\nn,
\end{equation}
with
\begin{equation}
 H_{\sss S} = {\Tr} \left\{ 3 \left(Y_{\sss U}^\dagger Y_{\sss U}\right)^2
 + 3 \left(Y_{\sss D}^\dagger Y_{\sss D}\right)^2 +  
\left(Y_{\sss E}^\dagger Y_{\sss E}\right)^2\right\} \nn.
\end{equation}
The mass of the Standard Model Higgs boson is given 
by $M_H^2 = \lambda \sVEV{\phi_{WS}}^2$ and, for definiteness, we 
take $M_H = 115~\GeV$ at weak scale.

In order to run the renormalisation group
equations down to $1~\GeV$, we use the following initial values:
\begin{eqnarray}
U(1):\quad & g_1(M_Z) = 0.462 \nn,\quad & g_1(M_{\rm Planck}) = 0.614 \nn,\\
SU(2):\quad & g_2(M_Z) = 0.651 \nn,\quad & g_2(M_{\rm Planck}) = 0.504 \nn,\\
SU(3):\quad & g_3(M_Z) = 1.22  \nn,\quad & g_3(M_{\rm Planck}) = 0.491 \nn.
\end{eqnarray}

\section{Method of numerical calculation}\label{sec:FN}
\indent\ According to our philosophy -- at the Planck scale, all coupling 
constants are complex numbers of order unity -- we evaluate the 
product of mass-protecting Higgs VEVs required 
for each mass matrix element and provide it  
with a random complex number of order one as a factor. This means that we 
assume essentially maximal $CP$ violation in all sectors, including 
the neutrino sector. Since the exact values of 
all these coupling constants are not known, our model
is {\em only} able to provide its results 
order of magnitudewise.  

In this way, we simulate a long chain of fundamental Yukawa couplings 
and propagators making the transition corresponding to an 
effective Yukawa coupling in the Standard Model. In the numerical 
computation we then calculate the masses and mixing angles time 
after time, using different sets of random numbers and, in the 
end, we take the logarithmic average of the calculated quantities 
according to the following formula:
\begin{equation}
  \label{eq:avarage}
  \sVEV{m}=\exp\left(\sum_{i=1}^{N} \frac{\ln m_i}{N}\right) \nn. 
\end{equation}
Here $\sVEV{m}$ is what we take to be the prediction for one of the 
masses or mixing angles, $m_i$ is the result of the calculation
done with one set of random number combinations and $N$ is the total 
number of random number combinations used.
The calculations are done using a Monte Carlo method
by putting in for the order of unity couplings random complex 
numbers of order unity. Strictly speaking, we just put such 
random number factors on the different mass matrix elements.
We then interpret the matrix elements estimated as the 
products of a random number of order unity and the Higgs 
VEVs (suppression factors) as the 
running Yukawa couplings at the scale where the Higgs fields $W$, 
$T$, $\rho$, $\omega$ and $\chi$ roughly all have their VEVs 
namely about one order of magnitude below the Planck scale. We 
ignore running of couplings over this single order of magnitude as 
not important.
We then run down case after case of random numbers the couplings 
to get the observable quantities at the experimental scale, which 
we take to be $1~\GeV$ as convention so that our masses are running ones at 
$\mu=1~\GeV$. At first we get the different results for each 
set of random numbers, but then we average over {\em the logarithm} 
of these quantities and present the exponents of the average log's,
as ``logarithmic averages''.
In estimating the quality of our fits of the adjustable VEVs we 
define a quantity which we call the goodness of fit (\mbox{\rm g.o.f.}) 
reminiscent of the chi-square using again the logarithms
\begin{equation}
\label{eq:gof}
 \mbox{\rm g.o.f.}\equiv\sum \left[\ln \left(
\frac{\sVEV{m}}{m_{\rm exp}} \right) \right]^2 \nn.
\end{equation}

It should be kept in mind that, in our calculations, we 
use the renormalisation group 
$\beta$-functions to run Yukawa couplings down to the experimentally 
observable scale $1~\GeV$. This is because we 
took the charged fermion masses to be compared to ``measurements'' 
at the conventional scale of $1~\GeV$, except for the 
top quark. We used the top quark pole mass instead~\cite{pierre}:
\begin{equation}
M_t = m_t(M)\left(1+\frac{4}{3}\frac{\alpha_s(M)}{\pi}\right)\nn,
\end{equation}
where we set $M=180~\GeV$ as an input, for simplicity.

\subsection{Mass matrices}\label{sec:mass}
\indent\ With the system of quantum numbers in Table~\ref{Table1} 
one can easily evaluate, for a given mass matrix 
element, the numbers of Higgs field VEVs of the different types 
needed to perform the transition between the corresponding left- and 
right-handed Weyl fields. The results of calculating the products of 
Higgs fields needed, 
and thereby the order of magnitudes of the mass matrix elements 
in our model, are presented in the following mass matrices:

\noindent
the up-type quarks:
\begin{eqnarray}
M_{\sss U} \simeq \frac{\sVEV{(\phi_{\sss\rm WS})^\dagger}}{\sqrt{2}}
\hspace{-0.1cm}
\left(\!\begin{array}{ccc}
        (\omega^\dagger)^3 W^\dagger T^2
        & \omega \rho^\dagger W^\dagger T^2
        & \omega \rho^\dagger (W^\dagger)^2 T\\
        (\omega^\dagger)^4 \rho W^\dagger T^2
        &  W^\dagger T^2
        & (W^\dagger)^2 T\\
        (\omega^\dagger)^4 \rho
        & 1
        & W^\dagger T^\dagger
\end{array} \!\right)\label{M_U}
\end{eqnarray}  
\noindent
the down-type quarks:
\begin{eqnarray}
M_{\sss D} \simeq \frac{\sVEV{\phi_{\sss\rm WS}}}
{\sqrt{2}}\hspace{-0.1cm}
\left (\!\begin{array}{ccc}
        \omega^3 W (T^\dagger)^2
      & \omega \rho^\dagger W (T^\dagger)^2
      & \omega \rho^\dagger T^3 \\
        \omega^2 \rho W (T^\dagger)^2
      & W (T^\dagger)^2
      & T^3 \\
        \omega^2 \rho W^2 (T^\dagger)^4
      & W^2 (T^\dagger)^4
      & W T
                        \end{array} \!\right) \label{M_D}
\end{eqnarray}
\noindent %
the charged leptons:
\begin{eqnarray}        
M_{\sss E} \simeq \frac{\sVEV{\phi_{\sss\rm WS}}}
{\sqrt{2}}\hspace{-0.1cm}
\left(\hspace{-0.1 cm}\begin{array}{ccc}
    \omega^3 W (T^\dagger)^2
  & (\omega^\dagger)^3 \rho^3 W (T^\dagger)^2 
  & (\omega^\dagger)^3 \rho^3 W T^4 \chi \\
    \omega^6 (\rho^\dagger)^3  W (T^\dagger)^2 
  &   W (T^\dagger)^2 
  &  W T^4 \chi\\
    \omega^6 (\rho^\dagger)^3  (W^\dagger)^2 T^4 
  & (W^\dagger)^2 T^4
  & WT
\end{array} \hspace{-0.1cm}\right) \label{M_E}
\end{eqnarray}
\noindent
the Dirac neutrinos:
\begin{eqnarray}
M^D_\nu \simeq \frac{\sVEV{(\phi_{\sss\rm WS})^\dagger}}{\sqrt{2}}
\hspace{-0.1cm}
\left(\hspace{-0.1cm}\begin{array}{ccc}
        (\omega^\dagger)^3 W^\dagger T^2
        & (\omega^\dagger)^3 \rho^3 W^\dagger T^2
        & (\omega^\dagger)^3 \rho^3 W^\dagger  T^2 \chi\\
        (\rho^\dagger)^3 W^\dagger T^2
        &  W^\dagger T^2
        & W^\dagger T^2 \chi\\
        (\rho^\dagger)^3 W^\dagger T^\dagger \chi^\dagger
        &  W^\dagger T^\dagger \chi^\dagger
        & W^\dagger T^\dagger
\end{array} \hspace{-0.1 cm}\right)\label{Mdirac}
\end{eqnarray} 
\noindent %
and the Majorana (right-handed) neutrinos:
\begin{eqnarray}    
M_R \simeq \sVEV{\phi_{\sss\rm B-L}}\hspace{-0.1cm}
\left (\hspace{-0.1 cm}\begin{array}{ccc}
(\rho^\dagger)^6 \chi^\dagger
& (\rho^\dagger)^3 \chi^\dagger /2
& (\rho^\dagger)^3/2  \\
(\rho^\dagger)^3 \chi^\dagger /2
& \chi^\dagger & 1 \\
(\rho^\dagger)^3/2 & 1 & \chi
\end{array} \hspace{-0.1 cm}\right ) \label{Majorana}
\end{eqnarray}       

In order to get the true model matrix elements, one must imagine 
that each matrix element is provided with an order of unity factor, 
which is unknown within our system of assumptions and which, as 
described above, is taken in our calculation as a complex random 
number, later to be logarithmically averaged over as in 
eq.~(\ref{eq:avarage}).

The off-diagonal elements of the right-handed neutrino mass matrix
(eq.~\ref{Majorana}) are divided by a factor 2 because the 
symmetric (Majorana) mass matrix gives rise to the same off-diagonal term 
twice, $\ie$, we avoid the overcounting of the corresponding Feynmann 
diagrams. However, the element which couples with only one Higgs field
should not be multiplied by an extra factor $1/2$.

Note that the quantum numbers of our $6$ Higgs fields
are not totally independent. In fact there is a linear relation
between the quantum numbers of the three Higgs fields $W$, 
$T$ and $\chi$:
\begin{equation}
 \vec{Q}_\chi= 3\,\vec{Q}_W - 9\,\vec{Q}_T  \nn,
\end{equation}
where the 6 components of the charge vector $\vec{Q}$ correspond 
to the 6 columns of Table~\ref{Table1}.
Thus the Higgs field combinations needed for a given transition are not
unique, and the largest contribution has to be selected for each matrix 
element in the above mass matrices. 

{}Furthermore, we should mention that we do \underline{not} assume 
Supersymmetry, so that we can make use of an expectation value of a 
Higgs field and of the Hermitian conjugate one without any 
restriction; they are numerically equal. If we supersymmetrised 
the model we would need to double the Higgs fields and we 
would loose predictive power, because we would get more parameters.

\subsection{Renormalisation group effect on Dirac neutrino Yukawa coupling}
\label{sec:randirac}
\indent\ 
In the previous work~\cite{FNT} we ignored the running of the Yukawa 
couplings 
for the neutrino sector from the Planck scale down to the see-saw 
scale and only used the running of the effective dimension five 
operator (eq.~\ref{eq:remeff}),
but in the  present article we use also the running of the Dirac 
Yukawa couplings for the neutrinos above the see-saw scale according 
to eq.~(\ref{Diracyukawa}). By putting in the various values we may estimate 
that the most important term 
is the top-Yukawa containing term, $Y_S$, which makes the overall size of the 
neutrino Dirac Yukawa matrix increase towards the ultraviolet. 
Since our predictions are {\it a priori} made at the Planck scale, this 
means that inclusion of this running makes the predictions for these 
Yukawa couplings a bit -- about a factor $\sqrt{2}$ or less -- smaller 
at lower energy scale.

\section{Results for the quantities of quarks and leptons}\label{sec:results}
\indent\ Using the three charged quark-lepton mass matrices and the effective 
neutrino mass matrix together with the renormalisation group 
equations we made a fit to all the fermion quantities 
in Table~\ref{convbestfit} varying just $6$ Higgs 
fields VEVs. We averaged over $N=10,000$ complex order unity random 
number combinations (see eq.~\ref{eq:avarage}). These complex numbers
are chosen to be the exponential of a number picked from a Gaussian 
distribution, with mean value zero and standard deviation one, 
multiplied by a random phase factor that has smoothly distributed 
phase. We varied the $6$ free 
parameters and found the best fit, corresponding to the lowest value 
for the quantity \mbox{\rm g.o.f.} defined in eq.~(\ref{eq:gof}), with the 
following values for the VEVs:
\begin{eqnarray} 
\label{eq:VEVS} 
&&\sVEV{\phi_{\sss WS}}= 246~\GeV\nn,  
\nn\sVEV{\phi_{\sss B-L}}=1.23\times10^{10}~\GeV\nn, 
\nn\sVEV{\omega}=0.245\nn,\nonumber\\
&&\nn\sVEV{\rho}=0.256\nn,\nn\sVEV{W}=0.143\nn,
\nn\sVEV{T}=0.0742\nn,\nn\sVEV{\chi}=0.0408\nn,
\end{eqnarray}
where, except for the Weinberg-Salam Higgs field and 
$\sVEV{\phi_{\sss B-L}}$, the VEVs are expressed in Planck units. 
Hereby we have considered that the Weinberg-Salam Higgs field VEV is 
already fixed by the Fermi constant.
The results of the best fit, with the VEVs in eq.~(\ref{eq:VEVS}), 
are shown in Table~\ref{convbestfit} and the fit has  
$\mbox{\rm g.o.f.}=3.38$ (see the definition in eq.~\ref{eq:gof}). 

\begin{table}[!b]
\begin{displaymath}
\begin{array}{|c|c|c|}
\hline\hline
 & {\rm Fitted} & {\rm Experimental} \\ \hline
m_u & 5.2~\MeV & 4~\MeV \\
m_d & 5.0~\MeV & 9~\MeV \\
m_e & 1.1~\MeV & 0.5~\MeV \\
m_c & 0.70~\GeV & 1.4~\GeV \\
m_s & 340~\MeV & 200~\MeV \\
m_{\mu} & 81~\MeV & 105~\MeV \\
M_t & 208~\GeV & 180~\GeV \\
m_b & 7.4~\GeV & 6.3~\GeV \\
m_{\tau} & 1.11~\GeV & 1.78~\GeV \\
V_{us} & 0.10 & 0.22 \\
V_{cb} & 0.024 & 0.041 \\
V_{ub} & 0.0025 & 0.0035 \\ \hline
\Delta m^2_{\odot} & 9.0 \times 10^{-5}~\eV^2 &  4.5 \times 10^{-5}~\eV^2 \\
\Delta m^2_{\rm atm} & 1.8 \times 10^{-3}~\eV^2 &  3.0 \times 10^{-3}~\eV^2\\
\tan^2\theta_{\odot} &0.23 & 0.35\\
\tan^2\theta_{\rm atm}& 0.83 & 1.0\\
\tan^2\theta_{\rm chooz}  & 3.3 \times 10^{-2} & \sleq~2.6 \times 10^{-2}\\
\hline\hline
\mbox{\rm g.o.f.} &  3.38 & - \\
\hline\hline
\end{array}
\end{displaymath}
\caption{Best fit to conventional experimental data.
All masses are running
masses at $1~\GeV$ except the top quark mass which is the pole mass.
Note that we use the square roots of the neutrino data in this 
Table, as the fitted neutrino mass and mixing parameters 
$\sVEV{m}$, in our goodness of fit ($\mbox{\rm g.o.f.}$) definition, 
eq.~(\ref{eq:gof}).}
\label{convbestfit}
\end{table}
We have $11=17 - 6$ degrees of freedom -- predictions -- leaving each of 
them with a logarithmic error of
$\sqrt{3.38/11}\simeq0.55$, $\ie$, we can fit {\rm all quantities} 
with a typical error of a factor 
$\exp\left(\sqrt{3.38/11}\right)=1.74$ of the experimental value.

Unlike in older versions of the model, the first and second 
family sub-matrix of $M_{\sss D}$ is now dominantly diagonal. In 
previous versions of the model this submatrix satisfied the 
order of magnitude factorisation condition $(M_{\sss D})_{12}\cdot
(M_{\sss D})_{21}$ $\approx$ $(M_{\sss D})_{11}\cdot(M_{\sss D})_{22}$; 
thus the down quark mass $m_d$ received two contributions 
(off-diagonal as well as diagonal) of the same order of magnitude 
as the up quark mass $m_u$. This extra off-diagonal contribution to 
$m_d$ of course improved the goodness of the fit to the masses of 
the first family, since phenomenologically $m_d \approx 2~m_u$. 
However, in the present version of the model with the $\omega$ 
and $\rho$ Higgs fields, the off-diagonal element
$(M_{\sss D})_{21}$ becomes smaller and we are left with a full 
order of magnitude degeneracy of the first family masses, even 
including the down quark. Our best fit 
values for the charm quark mass $m_c$ 
and the Cabibbo angle $V_{us}$ are smaller than in our 
previous fits to the charged fermion masses~\cite{NT,FF}. 
Nonetheless, as mentioned above, our present best fit agrees 
with the experimental data within the theoretically expected 
uncertainty of about $64\%$ and is, therefore, as well as can 
expect from an order of magnitude fit.  

\subsection{Neutrino oscillation parameters}
\indent\ The charge current interactions 
from the Sudbury Neutrino Observatory (SNO)~\cite{SNO} have 
provided an important signal confirming the 
existence of the solar neutrino 
mass~\cite{chlorine,sage,gallex,gno,SK8B}: SNO 
detected a flux of $\nu_\mu$ neutrino or $\nu_\tau$ neutrino
among solar neutrinos after
traveling from the core of the Sun to the Earth. 
Combination of the SNO results with previous measurements from 
other experiments confirms the standard solar 
model~\cite{Bahcall}, whose predictions of the total 
flux of active $^8$B neutrinos in the Sun agree with the SNO and 
Super-Kamiokande~\cite{SK8B} data.
{}Furthermore, the measurement of the $^8$B and $hep$ 
solar neutrino fluxes shows no significant energy dependence 
of the electron neutrino survival probability in the
Super-Kamiokande and SNO energy ranges. 
In fact, global 
analyses~\cite{fogli,cc1,goswami,smirnov} of solar 
neutrino data, including the first SNO results and the day-night 
effect~\cite{SKDN}, which disfavoured the SMA-MSW solution 
at the $95\%$ C.L., have confirmed that the LMA-MSW solution 
gives the best fit to the data and that the SMA-MSW solution 
is very strongly
disfavoured and only accepted at the $3\sigma$ level. 

{}Furthermore, the combination of
the results from atmospheric neutrino experiments~\cite{SK}
and the CHOOZ reactor experiment~\cite{CHOOZ} constrains 
the first- and third-generation mixing angle to be small,
$\ie$ the $3 \sigma$ upper bound is given by 
$\tan^2\theta_{\rm chooz}~\sleq~0.06$. This limit was 
obtained from a three flavour neutrino analysis 
(in the five dimensional parameter space -- 
$\theta_\odot$, $\theta_{\rm chooz}$, $\theta_{\rm atm}$,
$\Delta m^2_{\odot}$ and $\Delta m^2_{\rm atm}$), 
using all the solar and atmospheric neutrino data and
based on the assumption that
neutrino masses have a hierarchical structure, $\ie$ 
$\Delta m^2_{\odot}\ll\Delta m^2_{\rm atm}$~\cite{cc2}.

In Table~\ref{convbestfit} presented solar neutrino data,
however, come from a global 
two flavour analysis, which means that the first- and 
third-generation mixing angle is essentially put 
equal to zero, $\ie$, the dependence of $\theta_{13}$
on the solar neutrino parameters have been ignored. In 
principle, of course, we have to fit to neutrino parameters 
from a three flavour analysis.

The global three flavour analyses have been 
done by several authors~\cite{cc2,Valle,fogli2,serguey1} 
and they showed a significant 
influence of the non-zero CHOOZ angle on the solar neutrino
mass squared difference and mixing angle. In~\cite{serguey1} 
the relatively large solar neutrino mass squared difference lying 
in the LMA-MSW region (with the condition 
$\Delta m^2_{\odot}\sgeq10^{-4}~\eV^2$), 
the solar mixing angle and the CHOOZ reactor experiment data
were analysed using the three flavour analysis method. Their works
tell us that if the CHOOZ angle
becomes bigger than zero then the solar mixing angle becomes
smaller. This effect is more significant for the larger 
$\Delta m^2_{\odot}$ values. Because of this correlation,
our fit to the neutrino data should be somewhat better than 
that suggested by the \mbox{\rm g.o.f.} value; even including 
the CHOOZ angle our neutrino fit is extremely good.

Experimental results on the values of neutrino mixing angles 
are usually presented in terms of the function $\sin^22\theta$ 
rather than $\tan^2\theta$ (which, contrary to $\sin^22\theta$, 
does not have a maximum at $\theta=\pi/4$ and thus still varies 
in this region).
Transforming from $\tan^2\theta$ variables to $\sin^22\theta$ 
variables, our predictions for the neutrino mixing angles become:
\begin{eqnarray}
  \label{eq:sintan}
 \sin^22\theta_{\odot} &\!=\!& 0.61\nn,\\
 \sin^22\theta_{\rm atm} &\!=\!& 0.99\nn, \\
 \sin^22\theta_{\rm chooz} &\!=\!& 0.12\nn.
\end{eqnarray}  
We also give here our predicted hierarchical neutrino mass 
spectrum:
\begin{eqnarray}
m_1 &\!=\!& 9.8\times10^{-4}~~\eV\nn, 
\label{eq:neutrinomass1}\\
m_2 &\!=\!& 9.6\times10^{-3}~~\eV\nn, 
\label{eq:neutrinomass2}\\
m_3 &\!=\!& 4.4 \times10^{-2}~~\eV\nn.
\label{eq:neutrinomass3} 
\end{eqnarray}

Compared to the experimental data these predictions are 
excellent: all of our order of magnitude neutrino predictions lie 
inside the $95\%$ C.L. border determined from phenomenological fits 
to the neutrino data, even including the CHOOZ upper bound.
On the other hand, our prediction of the solar mass squared 
difference is a factor of $2$ larger than the global fit
data even though the prediction is inside of the LMA-MSW region, 
giving a contribution to our goodness of fit $\approx 0.12$. 


Our CHOOZ angle also turns out to be 
about a factor of $\sqrt{2}$ larger than the experimental limit at 
$90\%$ C.L., corresponding to another contribution of \mbox{\rm g.o.f.} 
$\approx 0.014$. In summary our predictions for the neutrino sector 
agree extremely well with the data, giving a contribution of only 
$0.25$ to \mbox{\rm g.o.f.} while the charged fermion sector contributes 
$3.13$ to \mbox{\rm g.o.f.}

Note that the value of $\sVEV{\phi_{B-L}}$ presented in 
Table~\ref{convbestfit} does not fit very well the atmospheric
neutrino mass squared difference, being about factor $\sqrt{2}$ less than 
the best fit reported by Super-Kamiokande collaboration~\cite{SK}.
That means that our model predicts ``relative'' degenerated mass spectra in 
the neutrino sector. However, if we force it to fit the 
value of mass squared difference as 
$\Delta m^2_{\rm atm}=2.5\times 10^{-3}~\eV^2$
and just arrange the scale, $\sVEV{\phi_{B-L}}$, of the mass squared 
differences -- it will not significantly influence the
ratio of mass squared 
differences -- thus we get $\Delta m^2_{\odot}=1.2\times 10^{-4}~\eV^2$
which lays still in the allowed region of the global fit of 
the solar neutrino analysis:
\begin{eqnarray}
  \label{eq:corrBL}
  \sVEV{\phi_{B-L}} \Big|_{~\rm corrected} &=& 1.0\times 10^{10}~\GeV \nn,\\
\label{eq:corratm}
  \Delta m^2_{\rm atm} \Big|_{~\rm corrected}&=& 2.5\times 10^{-3}~\eV^2\nn,\\
\label{eq:corrsolar}
  \Delta m^2_{\odot} \Big|_{~\rm corrected} &=&1.2\times 10^{-4}~\eV^2 \nn.
\end{eqnarray}
and the CHOOZ angle does not change 
$\tan^2\theta_{\rm chooz}=3.3\times10^{-2}$ in this small 
deviation of the see-saw scale.

\subsection{$CP$ violation}
\indent\ Since we have taken our random couplings to be -- 
whenever allowed -- \underline{complex} we have order of 
unity or essentially maximal $CP$ violation so that a unitary 
triangle with angles of order one is a success of our 
model. After our fitting of masses and of mixings 
we can simply predict order of magnitudewise the $CP$ violation in 
$\eg$ $K^0\! - \!\bar{K^0}$ decay or in CKM and MNS 
mixing matrices in general. 

The Jarlskog area $J_{\sss CP}$ provides a measure of the 
amount of $CP$ violation in the quark sector~\cite{cecilia} and, 
in the approximation of setting cosines of mixing angles to unity, 
is just twice the area of the unitarity triangle:
\begin{equation}
  \label{eq:jarkskog}
  J_{\sss CP}=V_{us}\,V_{cb}\,V_{ub}\,\sin \delta \,,
\end{equation}
where $\delta$ is the $CP$ violation phase in the CKM matrix.
In our model the quark mass matrix elements have random phases, 
so we expect $\delta$ (and also the three angles $\alpha$, 
$\beta$ and $\gamma$ of the unitarity triangle) to be of 
order unity and, taking an average value of 
$\abs{\sin\delta}\approx 1/2$, the area of the 
unitarity triangle becomes
\begin{equation}
  \label{eq:jarkskog*0.5}
  J_{\sss CP}\approx \frac{1}{2}\,V_{us}\,V_{cb}\,V_{ub}\,.
\end{equation}
Using the best fit values for the CKM elements from 
Table~\ref{convbestfit}, we predict 
$J_{\sss CP} \approx 3.0 \times10^{-6}$ to be compared with 
the experimental value $(2-3.5)\times10^{-5}$. 
Since our result for the Jarlskog area  
is the product of four quantities, we do not expect the 
usual $\pm64\%$ logarithmic uncertainty but rather 
$\pm\sqrt{4}\cdot64\%=128\%$ logarithmic
uncertainty. This means our result deviates from the 
experimental value by 
$\ln (\frac{2.7 \times 10^{-5}}{3.0 \times 10^{-6}})/1.28$ = 1.7 
``standard deviations''. 

The Jarlskog area has been calculated from the best fit parameters in 
Table~\ref{convbestfit}, it is also possible to calculate them directly while 
making the fit. So we have calculated $J_{\sss CP}$ 
for $N=10,000$ complex order unity random 
number combinations. Then we took the logarithmic average 
of these $10,000$ samples of $J_{\sss CP}$
and obtained the following result:
\begin{eqnarray}
  \label{eq:jcpabsm}
   J_{\sss CP}&=& 3.0\times 10^{-6} \nn,
\end{eqnarray}  
in good agreement with the values given above.

\subsection{Neutrinoless double beta decay}
\indent\ Another prediction, which can also be made from this model, is 
the electron ``effective Majorana mass'' -- the parameter in  
neutrinoless beta decay -- defined by: 
\begin{equation}
\label{eq:mmajeff}
\abs{\sVEV{m}} \equiv \abs{\sum_{i=1}^{3} U_{e i}^2 \, m_i} \nn,
\end{equation}
where $m_i$ are the masses of the neutrinos $\nu_i$ 
and $U_{e i}$ are the MNS mixing matrix elements for the 
electron flavour to the mass eigenstates $i$. We can 
substitute values for the neutrino masses $m_i$ from 
eqs.~(\ref{eq:neutrinomass1}-\ref{eq:neutrinomass3}) and for the 
fitted neutrino mixing angles from Table~\ref{convbestfit} into 
the left hand side of eq.~(\ref{eq:mmajeff}). 
As already mentioned, the $CP$ violating phases in the MNS mixing 
matrix are essentially random in our model. So we combine the 
three terms in eq.~(\ref{eq:mmajeff}) by taking the square root of 
the sum of the modulus squared of each term, which gives
our prediction:
\begin{equation}
  \label{eq:meffresult}
  \abs{\sVEV{m}} \approx 2.5\times 10^{-3}~~\mbox{\rm eV}\nn.
\end{equation}

In the same way as being calculated the Jarlskog area we
can compute using $N=10,000$ complex order unity random 
number combinations to get the $\abs{\sVEV{m}}$. Then we 
took the logarithmic average 
of these $10,000$ samples of $\abs{\sVEV{m}}$ as 
usual: 
\begin{eqnarray}
    \abs{\sVEV{m}}&=& 3.4\times 10^{-3}~~\mbox{\rm eV}\nn.
\end{eqnarray}  
This result does not agree with the central value of recent 
result -- ``evidence'' -- from the
Heidelberg-Moscow collaboration~\cite{evidence}.

\section{Baryogenesis via lepton number violation} 
\indent\ Now we have a good model which predicts orders of magnitude 
for all the Yukawa couplings including the see-saw particles,
so it is natural to ask ourselves whether this model can predict
also the right amount of $Y_B$ -- the ratio of the 
baryon numbers density relative to the entropy density -- 
using the Fukugita-Yanagida mechanism~\cite{FY}. According to 
this mechanism the decays of the 
right-handed neutrinos by $CP$ violating couplings lead 
to an excess of the $B-L$
charge (baryon number minus lepton number), the relative excess
in the decay from Majorana neutrino generation number $i$ 
being called $\epsilon_i$. This 
excess is then immediately -- and continuously back and forth -- being
converted partially to a baryon number excess, although it starts out 
as being a lepton number $L$ asymmetry, since the right-handed 
neutrinos decay into leptons and Weinberg-Salam Higgs particles.   

It is a complicated discussion to estimate to what extend the 
$B-L$ asymmetry
is washed-out later in the cosmological development, but 
in our approximation below we agree within a factor $3$ 
with the baryon number excess left to fit the 
Big Bang development at the stage of formation of the light elements 
primordially (nuclearsynthesis). The ``experimental'' data
of the ratio of baryon number density to the entropy density
is obtained by recent measurement of cosmic microwave background
radiation~\cite{dibari}:  
\begin{equation}
  \label{eq:YBexp}
  Y_B~\Big|_{\rm exp}=\left(8.5{+1.5\atop-1.0}\right)\times10^{-11}\nn.
\end{equation}

Recently, baryogenesis calculations had been done by several
authors~\cite{BACMF} using different models based on
this scenario, and they used ``usual'' range of $Y_B$:
\begin{equation}
  \label{eq:YBexpnor}
  Y_B~\Big|_{\rm exp}=\left(1.7-8.1\right)\times10^{-11}\nn.
\end{equation}

In the following subsection, we review
briefly the already known Fukugita-Yanagida mechanism.

\subsection{Fukugita-Yanagida scenario for lepton number production}
\indent\ The $SU(2)$ instantons~\cite{'tHooft}, rather say, 
Sphaleron~\cite{sphaleron} guaranteed the rapid
exchanges of the minus baryon number and lepton number in which 
though $B-L$ is conserved in the time of Big Bang, even when the 
temperature was above the weak scale. 
In fact the three right-handed neutrinos decay in the $B-L$ 
violating way, at the temperature about the see-saw mass 
scales. This means that the baryon number
is violated. From our assumption -- all fundamental coupling 
constants are of order of unity and we treat them as complex random numbers 
at the Planck scale -- it is clear that the model has not 
only $C$ violation but also $CP$ violation. Out-of-equilibrium 
condition comes about during the Hubble expansion 
due to the excess of the three type of the right-handed Majorana 
neutrinos caused by their masses. 

These statements lead to that 
all Sakharov conditions~\cite{sakharov}
are fulfilled in our model: {\rm (1)} baryon number violation, 
{\rm (2)} $C$ and $CP$ violation, {\rm (3)} departure 
from thermal equilibrium. 

\subsection{Entropy density in cosmology}
\indent\ In order to investigate the quantity, $Y_B$, we 
need the expression for the entropy of Planck radiation which 
is given by
\begin{equation}
  \label{eq:entropy}
  s_i = \frac{2 \pi^2 \,g_{*\,i}}{45} \,T^3 \nn,
\end{equation}%
where $g_{*\,i}$ is the total number of effectively massless degrees of
freedom of the plasma at the temperature of the right-handed
neutrino, and the index $i$ denotes the number of
copiously existing right-handed neutrinos at the time in 
question. For the estimate of the excess of baryon number 
coming from a certain right-handed neutrino $i$, we should 
compare to the entropy density at the temperature being approximately 
equal to the mass of the corresponding right-handed neutrino.

The $g_{*\,i}$ are obtained as follows: There are $14$ bosons and 
$45$ well-known Weyl fermions plus
$i$ Majorana particles when the temperature is high enough
compared to $10^2~\GeV$:
\begin{eqnarray}
  \label{eq:gstarAGUT}
g_{*\,i} \!\! &=&\!\! \sum_{j={\rm bosons}}g_{\,j}\left(\frac{T_j}{T}\right)^4+
\frac{7}{8}\sum_{j={\rm fermions}}g_{\,j}\left(\frac{T_j}{T}\right)^4 %
\nonumber\\
\!\! &=&\!\! 
\underbrace{28\,+ \,\frac{7}{8}\cdot90}_{\textrm{Standrad Model}}\,
+ \underbrace{\frac{7}{4}\cdot i}_{\textrm{see-saw particles}}\,
\nonumber\\
&=& \left\{ \begin{array}{r@{\quad:\quad}l}
            108.5 & i=1\\110.25 & i=2 \\ 112 & i=3
            \end{array}\right.\nn,
\end{eqnarray}
here $T_j$ denotes the effective temperature of any species $j$.
When we have coupling as at the stage discussed between all the 
particles $T_j=T$. 

\subsection{$CP$ violation in decays of the Majorana neutrinos}
\label{sec:majodecay}
\indent
A right-handed neutrino, $N_{{\sss R}\, i}$, decays 
into a Weinberg-Salam Higgs 
particle and a left-handed lepton or into the $CP$ 
conjugate channel: These two channels have different 
lepton numbers $\pm1$. If there were no $CP$ violation 
they would have the same branching ratio. However, all Yukawa
couplings are \underline{complex} order unity random numbers,
therefore the partial widths 
do not have to be equal in the next-to-leading-order of 
perturbation theory: $CP$ violation in the decay of right-handed 
neutrinos.

Defining the measure $\epsilon_i$ for the $CP$ violation 
\begin{equation}
  \label{eq:epsilonCP}
 \epsilon_i \equiv\frac{\sum_{\alpha,\beta}\Gamma(N_{{\sss R}\, i} \to \ell^\alpha\phi_{\sss WS}^\beta)-\sum_{\alpha,\beta}\Gamma(N_{{\sss R}\, i}\to \bar\ell^\alpha \phi_{\sss WS}^{\beta \dagger})}{\sum_{\alpha,\beta}\Gamma(N_{{\sss R}\, i}
\to \ell^\alpha\phi_{\sss WS}^\beta) + \sum_{\alpha,\beta}\Gamma(N_{{\sss R}\, i}\to\bar\ell^\alpha \phi_{\sss WS}^{\beta \dagger})}\nn, 
\end{equation}
where $\Gamma$ are $N_{{\sss R}\, i}$ decay rates (in the $N_{{\sss R}\, i}$ 
rest frame), summed over the
neutral and charged leptons (and Weinberg-Salam Higgs fields) 
which appear as final states in the $N_{{\sss R}\, i}$ decays 
one sees that the 
excess of leptons over anti-leptons produced in the decay
of one $N_{{\sss R}\, i}$ is just $\epsilon_i$.
Now we shall calculate $\epsilon_i$ in perturbation theory:
The total decay rate at the tree level (Fig.~\ref{Majodiagram}$(a)$) is 
given by
\begin{equation}
  \label{eq:LOCP}
  \Gamma_{N_i}=\Gamma_{N_i\ell}+\Gamma_{N_i\bar\ell}
={((\Tilde{M_\nu^D})^\dagger \Tilde{M_\nu^D)}_{ii}\over 
4\pi \sVEV{\phi_{\sss WS}}^2}\,M_i \nn,
\end{equation}%
where $\Tilde{M_\nu^D}$ can be expressed through the 
unitary matrix diagonalising the right-handed neutrino 
mass matrix $V_R$:
\begin{eqnarray}
  \label{eq:tildemd}
 \Tilde{M_\nu^D} \!&\equiv&\! M_\nu^D\,V_R \nn,\\
V_R^\dagger \,M_R\,M_R^\dagger\, V_R \!&=&\! 
{\rm diag} \left(\,M^2_1, M^2_2, M^2_3\,\right) \nn.
\end{eqnarray}
\begin{figure}[t!]
\vspace{5mm}
  \begin{center}
\unitlength=1mm
\begin{fmffile}{fmffig}
\newenvironment{NHf}
    {\begin{fmfgraph*}(35,30)
        \fmfleft{N}\fmfright{fb,f}
        \fmflabel{$\phi_{\sss WS}$}{fb}
        \fmflabel{$N_i$}{N}\fmflabel{$\ell$}{f}}
   {\end{fmfgraph*}}
  \begin{NHf}
    \fmf{plain}{N,v}\fmf{fermion}{v,f}
    \fmf{dashes}{v,fb}
    \fmfdot{v}
\fmflabel{\raisebox{-50mm}{$\!\!\!\!\!\!\!\!\!\!\!\!(a)$}}{v}
 \end{NHf}
\qquad \qquad
 \begin{NHf}
   \fmf{plain}{N,v1}\fmf{plain,tension=.65,label=$N_j$,l.side=left}{v2,v3}
   \fmf{dashes,right,tension=.25}{v1,v2}
   \fmf{fermion,right,tension=.25}{v2,v1}
   \fmf{fermion,tension=.35}{v3,f}
   \fmf{dashes,tension=.35}{v3,fb}
   \fmfdotn{v}{3}
\fmflabel{\raisebox{-50mm}{$\!\!\!\!\!\!\!\!\!\!\!\!(b)$}}{v3}
 \end{NHf}
\qquad \qquad
 \begin{NHf}
   \fmf{plain,tension=1.4}{N,Nv}
   \fmf{fermion,tension=.75}{fbv,Nv}
   \fmf{dashes,tension=.75}{Nv,fv}
   \fmf{dashes}{fbv,fb}\fmf{fermion}{fv,f}
   \fmffreeze
   \fmf{plain,label=$N_j$,l.side=right}{fbv,fv}\fmfdot{Nv,fbv,fv}
\fmflabel{\raisebox{-14.5mm}{$\!\!\!\!\!\!\!\!\!(c)$}}{fbv}
 \end{NHf}
\end{fmffile}
\vspace{12mm}
\caption{Tree level $(a)$, self-energy $(b)$ and vertex $(c)$ diagrams contributing to heavy Majorana neutrino decays.}
\label{Majodiagram}
\end{center}
\end{figure}
The $CP$ violation in the Majorana neutrino decays, 
$\epsilon_i$, arises when the effects of loop are 
taken into account, and at the one-loop level, the $CP$
asymmetry comes both from the wave function renormalisation 
(Fig.~\ref{Majodiagram}$(b)$) and from the vertex 
(Fig.~\ref{Majodiagram}$(c)$)~\cite{Luty,BuPlu,CRV}:
\begin{equation}
\label{eq:CPepsilon}
\epsilon_i = \frac{1}{4 \pi \sVEV{\phi_{\sss WS}}^2 ((\Tilde{M_\nu^D})^{\dagger} \Tilde{M_\nu^D})_{ii}}\sum_{j\not= i} {\rm Im}[((\Tilde{M_\nu^D})^{\dagger} \Tilde{M_\nu^D})^2_{ji}] \left[\, f \left( \frac{M_j^2}{M_i^2} \right) + g \left( \frac{M_j^2}{M_i^2} \right)\,\right] \nn,
\end{equation}
where the function, $f(x)$, comes from the one-loop vertex contribution and
the other function, $g(x)$, comes from the self-energy contribution.
These $\epsilon$'s can be calculated in perturbation theory 
only for differences between Majorana neutrino masses which 
are sufficiently large compare to their decay widths, $\ie$, the 
mass splittings satisfy the condition, 
$\abs{M_i-M_j}\gg\abs{\Gamma_i-\Gamma_j}$:
\begin{eqnarray}
f(x) &=& \sqrt{x} \left[1-(1+x) \ln \frac{1+x}{x}\right]\nn, \label{eq:vas1}\\
g(x) &=& \frac{\sqrt{x}}{1-x} \label{eq:vas2}\nn.
\end{eqnarray}
The function $f(x)$ has for large $x$ ``tricky'' cancellations, so
for numerical calculation we use following approximation in the region 
$x \gg 1$, $\ie$, hierarchical case in the right-handed sector:  
\begin{equation}
  \label{eq:approx}
  f(x)\simeq -\frac{1}{2 \, \sqrt{x}}\nn.
\end{equation}

\subsection{Off-diagonal dominant matrix elements in the 
see-saw neutrino mass matrix}
\label{sec:23domi}     
\indent\ The main modification in the model -- and not only in the 
calculation -- compared to the latest version is that we 
changed the quantum number assignment of the see-saw scale 
fitting Higgs field, $\phi_{B-L}$, so that it instead of 
giving directly and thus unsuppressed the $(3,3)$ mass matrix 
element, $\ie$,
a diagonal one, it gives in the present version unsuppressed size 
to the $(2,3)$ and $(3,2)$ mass matrix elements in this 
see-saw neutrino mass matrix (eq.~\ref{Majorana}). By having such 
off-diagonal dominance one expects a large mixing of the second and third 
generation at least in the see-saw mass matrix and expect 
to get easily a large CHOOZ angle $\tan^2\theta_{\rm chooz}$. 
However, as we have already presented in the section~\ref{sec:results}, 
we still do fit the CHOOZ angle limit satisfactorily.

The gain by having this off-diagonal dominance is, however, crucial 
for the baryon number production in the early Universe via 
the Fukugita-Yanagida mechanism: By having now the two 
heaviest see-saw neutrinos almost degenerate in mass 
the interaction between these two levels become much enhanced
and this causes the $CP$ violation as expressed by the $\epsilon_i$'s 
for the two heaviest Majorana neutrinos to be bigger, as is 
easily understood by noticing the factor $f(x)+g(x)$ occurring 
in the expression (eq.~\ref{eq:CPepsilon}),
which is in first approximation a suppression factor given by the 
ratio of the see-saw neutrino masses~\cite{API,FHY}. With the mass 
difference becoming 
small compared to the masses themselves the term $g(x)$ can even become 
big compared to unity, so that one might see the possibility of these
normally suppressing factors causing an enhancement. However, there are 
in the case with two such close Majorana neutrinos a strong cancellation 
between the amounts of $B-L$ produced in excess by these almost mass 
degenerate particles decaying. Nevertheless we can avoid significantly 
the suppression by having such degeneracy, therefore the 
baryon number production increases. 



\section{An approximation to the wash-out effect}\label{Kfactor}
\indent\ The excess $B-L$ produced at first gets partly 
destroyed by the same
or other $B-L$ violating processes. There are several 
processes/Feynmann diagrams that can lead to $B-L$ 
violation and thus wash-out an excess. Most important at 
the temperature scale of the right-handed neutrino masses 
are the resonance scattering of a Weinberg-Salam Higgs 
towards a left-handed lepton producing one of the Majorana 
neutrinos as a resonance. The rate of forming resonances this way 
is proportional to the width $\Gamma_i$ of the see-saw neutrino
functioning as resonance from detailed balance. Since the time scale involved 
is given by the Hubble expansion at the time of the temperature 
being equal to the mass of the see-saw neutrino, a crucial parameter 
for the wash-out effect via the resonance process is
\begin{equation}
  \label{eq:Kdrei}
K_i\equiv\frac{\Gamma_i}{2 H} \,\Big|_{ T=M_{i} } = \frac{M_{\rm
Planck}}{1.66 \sVEV{\phi_{\sss WS}}^2  8 \pi 
g_{*\,i}^{1/2}}\frac{((\Tilde{M_\nu^D})^{\dagger} 
\Tilde{M_\nu^D})_{ii}}{M_{i}} \qquad
(i=1, 2, 3)\nn, \end{equation}%
where $\Gamma_i$ is the width of the flavour $i$ Majorana neutrino,
$M_i$ is its mass and $g_{*\,i}$ is the number of degrees of freedom
at the temperature $M_i$.

As the time goes in early cosmology, the heaviest see-saw
neutrino goes out-of-thermal-equilibrium first and 
deploys its excess of $B-L$, then the next and so on. Thus in 
the hierarchical case -- but, our model is not this 
case in as far as the two heaviest see-saw neutrinos 
are approximately degenerate in mass -- the excess 
produced by the lightest see-saw neutrino will roughly only be 
washed out by itself but not by the heavier neutrinos, 
since the latter resonances are not reachable at the 
temperature at that time. However, the excess from the decay 
of the heavier see-saw neutrinos can be washed out by the 
lighter one(s), and if some are degenerate they may wash out
the products of each other. Now the flavours of the 
left-handed leptons produced in excess in the decays are not the
same for the three different see-saw neutrinos. They are, however, 
also not orthogonal states because they are mixing almost 
maximally. This means that wash-out due to one see-saw neutrino of 
the excess caused by another one is suppressed compared to 
what the wash-out rate would be when a see-saw neutrino washes 
out its {\em own} production of excess. We shall take into 
account such suppressions by constructing some effective
$K_i$ called ${K_{\sss\rm eff}}_i$ which should mean that value
of $K_i$ to use as if we had the excess of the ${\nu_{\sss R}}_i$ 
washed out by itself, so as to obtain in practice the effect 
of the wash-out due to the other see-saw neutrinos included.
The decay products of a given see-saw neutrino are found 
from the Yukawa couplings of this right-handed neutrino to 
a Weinberg-Salam Higgs and a left-handed lepton, and these
couplings are proportional to the $M_\nu^D$-mass matrix 
elements. We -- crudely -- shall estimate ${K_{\sss\rm eff}}_i$ 
expressions in terms of $K_i$'s  which are 
the parameters for the wash-out of the products of the 
right-handed see-saw neutrinos itself, where $i=1$ is the lightest
and three ($i=3$) the heaviest.

In order to estimate the effective $K$ factors we first 
introduce some normalized state vectors for the decay products:
\begin{equation}
\label{eq:braket}
\ket{ i }\equiv \frac{1}{\sqrt{\sum_{k=1}^{3} 
\abs{\,\left[\Tilde{M_\nu^D}(M_i)\right]_{k i}}^2}\,} \,
\left(\,\left[ \Tilde{M_\nu^D}(M_i)\right]_{1 i} 
\,, \left[\Tilde{M_\nu^D}(M_i)\right]_{2 i}
\,, \left[\Tilde{M_\nu^D}(M_i)\right]_{3 i}\,\right)\nn,
\end{equation}%
\noindent
and further define their overlap:
\begin{equation}
  \label{eq:betaij}
  \zeta_{i j} \equiv \abs{\braket{i | j}}^2 \nn.
\end{equation}
Then we may take an approximation for the effective $K$ factors:
\begin{eqnarray}
\label{eq:keff1}
{K_{\sss\rm eff}}_1 &=& K_1(M_1)\nn, \\
\label{eq:keff2}
{K_{\sss\rm eff}}_2 &=& K_2(M_2) + \zeta_{23}\,K_3(M_3) 
+ \zeta_{21}\,K_1(M_1)\nn,\\ 
\label{eq:keff3}
{K_{\sss\rm eff}}_3 &=& K_3(M_3) + \zeta_{32}\,K_2(M_2) + 
\zeta_{31}\,K_1(M_1) \nn.
\end{eqnarray}
Strictly speaking, one shall namely not imagine the excesses to
flow around as left-handed leptons only, because there is 
with high rate the weak instanton or Sphaleron activity 
converting $\eg$ lepton number into minus baryon 
number. Since, however, the instanton process at the 
same time convert one particle from every generation, there 
remains the same excess of one generation over the 
other even after the instanton-process. Thus we expect 
that the assumed suppressed wash-out effect from a 
different see-saw neutrino works anyway.
We have for simplicity ignored the scattering processes due to
exchange processes as candidates for wash-out effects, 
expecting that at the temperature scales relevant they will 
effectively be small due to cross-section going as the fourth power
of Dirac neutrino Yukawa couplings rather than the square as does
the resonance cross-section averaged over energy.

\subsection{The dilution factors} 
\indent\ When the ${K_{\sss\rm eff}}$'s are not small (not less than one), 
we have to 
take into account the dilution effects -- wash-out effects -- 
for the calculation. Therefore, we define the suppression 
factor $\kappa_i$ to be fraction of $B-L$ excess
produced by the right-handed neutrino number $i$ which 
survives. That is to say, 
we define $\kappa_i$ so that the resulting relative to entropy density, $s_i$,
baryon number density minus lepton number, $\ie$, $Y_{B-L}$ becomes
\begin{equation}
  \label{eq:finalformBL}
  Y_{B-L} \equiv \left|\,\sum_{i=1}^{3}\kappa_i\,\frac{\epsilon_i}{g_{*\,i}} %
\,\right|\nn.
\end{equation}
A good approximation for $\kappa_i$, the dilution factor,
is inferred from refs.~\cite{API,KT}:
\begin{eqnarray}
{K_{\sss\rm eff}}_i\,\sgeq 10^6:\qquad 
\kappa_i&=& (0.1\,{K_{\sss\rm eff}}_i)^{\frac{1}{2}} \exp\left[-\frac{4}{3}\,(0.1\,{K_{\sss\rm eff}}_i)^{\frac{1}{4}}\right]\nn,\\
10 \sleq\, {K_{\sss\rm eff}}_i \,\sleq 10^6:\qquad\kappa_i&=&\frac{0.3}{{K_{\sss\rm eff}}_i(\ln {K_{\sss\rm eff}}_i)^{\frac{3}{5}}}\nn,\\
\label{gl:Keinszehn}
1 \sleq\, {K_{\sss\rm eff}}_i\,\sleq 10:\qquad\kappa_i &=& \frac{1}{2 \,{K_{\sss\rm eff}}_i}\nn,
\\
\label{gl:Knulleins}
0 \sleq\, {K_{\sss\rm eff}}_i \,\sleq 1:\qquad\kappa_i &=& 1\nn.
\end{eqnarray}

Note that these dilution factors are not smooth; therefore
we used in numerical calculations instead of the 
eq.~(\ref{gl:Keinszehn}) and eq.~(\ref{gl:Knulleins}) the 
following interpolating redefined dilution factor 
in the range $0\sleq\,{K_{\sss\rm eff}}_i \,\sleq 10$:
\begin{equation}
  \label{eq:neudi}
0\sleq\,{K_{\sss\rm eff}}_i \,\sleq 10:\qquad
\kappa_i = \frac{1}{\sqrt{{4\, {K_{\sss\rm eff}}_i}^2 + 1}} \nn.
\end{equation}

\subsection{$B-L$ to baryon number conversion}
\indent\ We have presented all quantities to construct
$Y_{B-L}$ in foregoing sections, moreover, we should note 
here that due to the 
electroweak Sphaleron effect, 
the baryon number asymmetry $Y_B$ is related to the baryon number minus 
lepton number asymmetry $Y_{B-L}$~\cite{HT}: %
\begin{eqnarray}
\label{BLtoB}
  Y_B = \frac{8N_f+4N_H}{22N_f+13 N_H} \,Y_{B-L} \nn,
\end{eqnarray}
where $N_f$ is the number of generations and $N_H$ the number of Higgs
doublets. It turns out that the 
final form of the baryogenesis using eq.~(\ref{eq:finalformBL}) is: 
\begin{equation}
\label{eq:finalB}
  Y_{B} = \frac{28}{79}\,\left|\,\sum_{i=1}^{3}\kappa_i\,
\frac{\epsilon_i}{g_{*\,i}}\,\right|\nn.
\end{equation}

\section{Result of baryogenesis}
\indent\ The numerical calculation of baryogenesis will be presented 
in this section. All the Yukawa couplings -- VEVs of seven different
Higgs fields -- are obtained by fitting the fermion quantities 
(see section $6$). In order to get baryogenesis in Fukugita-Yanagida
scheme, we have to calculate the following informations, $\ie$, the 
see-saw neutrino masses, $K$ factors and $CP$ violation parameters 
($\epsilon_i$'s). At first, we present the see-saw neutrino masses then
${K_{\sss\rm eff}}_i$'s using $N=10,000$ random number 
combinations and logarithmic average:
\begin{eqnarray}
\label{eq:Rmajomass1}
  M_1&=& 2.1 \times 10^{5}~\GeV\nn,\\
\label{eq:Rmajomass2}
  M_2&=& 8.8 \times 10^{9}~\GeV\nn,\\
\label{eq:Rmajomass3}
  M_3&=& 9.9 \times 10^{9}~\GeV\nn,
\end{eqnarray}
and 
\begin{eqnarray}
  \label{eq:Keff123}
  {K_{\sss\rm eff}}_1&=& 31.6\nn,\\
  {K_{\sss\rm eff}}_2&=& 116.2\nn,\\
  {K_{\sss\rm eff}}_3&=& 114.7\nn.
\end{eqnarray}
The numerical results of our best fitting VEVs also 
gives the $\epsilon_i$'s using same method:
\begin{eqnarray}
  \label{eq:epsion1231}
  \abs{\epsilon_1}&=& 4.62\times 10^{-12}\nn,\\
  \label{eq:epsion1232}
  \abs{\epsilon_2}&=& 4.00\times 10^{-6}\nn,\\
  \label{eq:epsion1233}
  \abs{\epsilon_3}&=& 3.27\times 10^{-6}\nn.
\end{eqnarray}
The sign of $\epsilon_i$ is unpredictable due to the complex 
random number coefficients in mass matrices, therefore we are 
not in the position to say the sign of $\epsilon$'s. It turns 
out that we should calculate directly $Y_B$ instead of 
putting the results from eqs.~(\ref{eq:Rmajomass1})-(\ref{eq:Rmajomass3}) 
and eqs.~(\ref{eq:epsion1231})-(\ref{eq:epsion1233}) using 
the dilution factor formulae
into eq.~(\ref{eq:finalB}). Using the complex order unity random 
numbers being given by a Gaussian distribution we get after logarithmic
average
\begin{equation}
  \label{eq:YB}
  Y_B = 2.59{+17.0\atop-2.25}\times 10^{-11} \nn,
\end{equation}
hereby we estimate the uncertainty  
in the natural exponent according ref.~\cite{FF} 
to be $64~\%\cdot\sqrt{10}\approx 200~\%$. Dominantly the 
signs of $\epsilon_i$ are strongly correlated, therefore, we 
included in the numerical calculation using 
eq.~(\ref{eq:finalB}) the correlation 
of signs effects correctly. 

The other remark of the scale dependence for our 
prediction is on the baryogenesis calculation:
It is obvious that the quantities, $\epsilon_i$, are approximately 
independent of the scale $\sVEV{\phi_{B-L}}$ from 
the eq.~(\ref{eq:CPepsilon}). In fact these quantities 
are influenced by only the ratio of the different mass 
eigenstates of the right-handed neutrino, and in addition 
through the Dirac neutrino Yukawa 
couplings which are, of course, depended on the see-saw
scale via renormalisation group equations. However, a small 
deviation of the see-saw scale can not contribute 
significantly. The number of degrees of freedom, $g_{*{}i}$, 
does not depend on the see-saw scale. However, the dilution
factors, rather say, the ${K_{\sss\rm eff}}_i$ factors are 
strongly influenced by
the right-handed scale, $\ie$, inverse power of the right-handed neutrino 
masses (see eq.~\ref{eq:Kdrei}). Strictly speaking, the 
wash-out effect is proportional to the see-saw scale. Using 
the ``right'' value of the see-saw scale
which we mention in above (eq.~\ref{eq:corrBL}) we get the corresponding 
${K_{\sss\rm eff}}$'s as following
\begin{eqnarray}
\label{eq:coK1}
{K_{\sss\rm eff}}_1~\Big|_{\rm ~corrected} &=& 37.60\nn,\\
\label{eq:coK2}
{K_{\sss\rm eff}}_2~\Big|_{\rm ~corrected} &=& 138.8\nn,\\
\label{eq:coK3}
{K_{\sss\rm eff}}_3~\Big|_{\rm ~corrected} &=& 137.1\nn.
\end{eqnarray}
The corresponding result of baryon asymmetry is
\begin{equation}
  \label{eq:coYB}
  Y_B~\Big|_{\rm corrected} = 2.02{+13.3\atop-1.75}\times 10^{-11}\nn.
\end{equation}
This result is a bit smaller than the result in eq.~(\ref{eq:YB}), 
however, still in allowed region of the ``experimental'' data in 
eq.~(\ref{eq:YBexpnor}).

\section{Proton decay}
\indent\ It should be mentioned that our non-supersymmetric model
passes without any trouble the test of not predicting 
proton decay that should have been observed in present 
experiment. Indeed we have 
in our model desert, except for the see-saw neutrinos and 
associated Higgs, $\phi_{B-L}$, and gauge $U(1)_{B-L}$ fields,  
up to about one order of magnitude below the Planck 
scale. 

A reasonable suggestion is that the baryon decay 
causing bosons would have 
masses of the order of the scale around the VEVs of our 
Higgs fields, which typically are about a factor $10$ 
below the Planck scale (see eq.~(\ref{eq:VEVS})). We 
should therefore use such a boson with mass about 
$1.2\times 10^{18}~\GeV$. That is to say we could 
essentially use the formula for the life time of the 
proton $\tau_p$ in simple $SU(5)$ GUT with replaced 
mass of $X$-particle by $1.2\times 10^{18}~\GeV$. We could 
take for instance the expression for the $p\to\pi^0\,e^+$ 
channel~\cite{murayama}: 
\begin{equation}
  \label{eq:protonzerfall}
\tau_{(p\to\pi^0\,e^+)} =\frac{1}{\Gamma(p\to\pi^0\,e^+)}=8 \times 10^{34}~{\rm years}\,\cdot \,\left(\frac{0.015~\GeV^3}{\alpha_H}\right)^2\,\cdot
 \,\left(\frac{M_V}{10^{16}~\GeV}\right)^4 \nn,
\end{equation}
where $\alpha_H$ is the hadronic matrix element and $M_V$ 
the mass of the GUT gauge boson, and insert $1.2\times10^{18}~\GeV$.
Then we get crudely our prediction assuming the existence of bosons 
which break baryon number at this scale:
\begin{equation}
  \label{eq:protonleben}
  \tau_{(p\to\pi^0\,e^+)} \approx 10^{43}~{\rm years}\nn.
\end{equation}

However, non of the Higgs fields and gauge 
bosons, which we truly consider in our model would 
be able to cause baryon decay. In fact we have excluded 
by our assumption about our gauge group the possibility of a 
gauge boson that could have made transitions between 
quarks and leptons: we excluded, namely, gauge bosons unifying 
different, irreducible representations of the Standard Model, 
and of course -- it is well-known -- quarks and leptons 
belong to different irreducible representations. Concerning the 
possibility that our scalar fields (Higgs fields) considered be 
able to cause baryon decay, it should 
be remarked that since they are made to break our gauge group 
(eq.~\ref{eq:agut}) but to leave unbroken the Standard Model 
gauge group. Therefore, they must contain a singlet under the diagonal 
group in order not to break the diagonal $SU(3)$ -- QCD gauge group.
Then it follows that they must under this diagonal $SU(3)$ 
have triality $t=0$~(mod $1$) (see below eq.~(\ref{eq:mod})).
In turn, that means that these Higgs fields are not able to 
couple to neither two quarks nor to a pair of a quark and a 
lepton. That makes our scalar fields of no practical 
use for baryon decay since to look for \underline{three} 
quarks coming together to form one point particle at a 
scale of $10^{18}~\GeV$ is too much requirement. 


If we keep to truly 
postulated particles in our model we would have to 
wait for proton decay till we can get sensitive to it 
from Planck scale masses inserted for $M_V$. Taken this 
attitude of only using ``the everything'' that can be 
found at Planck scale will lead to a $10^{4}$ times 
longer proton life time prediction in our model:
\begin{equation}
  \label{eq:protonlebensehrlang}
  \tau_{(p\to\pi^0\,e^+)} \Big|_{\rm Planck} \approx 10^{47}~{\rm years}\nn.
\end{equation}
In any case there is no chance in foreseeable future to 
observe proton decay according to the suggestions of 
our model -- however presumably -- this is a point 
that could be modified without changing our model 
too drastically.

\section{Conclusions}
\label{sec:conclution}
\indent\ We have set up a model for calculating all the mass 
matrices for quarks and leptons -- here under especially the 
effective left-handed neutrino mass matrix which is obtained via 
the well-known see-saw mechanism using a set of three 
see-saw neutrinos, two of which are approximately degenerate in mass. We 
have fit all the masses and mixing angles within the accuracy 
which we theoretically expect to be $\pm64\%$ (on a logarithmic 
scale) for the quantities for instance quark and lepton masses which 
are essentially directly given by mass matrix elements. In fact 
even our worst deviations of our predictions form experiment are 
of these to mass matrices simply related quantities: we predict 
the charm quark mass and the 
Cabibbo angle a factor two lower than experiment, while we obtain the 
electron and the strange mass a factor two larger.
On the other hand, the squared quantities of neutrino 
oscillations presented in Table~\ref{convbestfit} do not even 
deviate a factor two. The $CP$ violating parameter --
Jarlskog triangle area --, $J_{\sss CP}$, and the baryon number 
relative to the entropy ratio, $Y_B$, arise as products 
and ratios of several mass matrix elements and thus have 
larger expected uncertainties, and turn out indeed to 
agree quite within the expected accuracy both being 
predicted to the low side of ``data''. In conclusion 
we fit all the Standard Model parameters related to 
the mass matrices within the expected order of 
magnitude accuracy, $\ie$, we fitted $9$ masses of 
quark and charged leptons,
$2$ mass squared differences for neutrinos, $5$ genuinely measured 
mixing angles, the Jarlskog triangle area and the baryon number to 
entropy density ratio. In addition, we managed to avoid violating 
the bounds for proton decay life time and neutrinoless 
double beta decay. Most interestingly we predicted the CHOOZ mixing 
angle -- the mixing angle of the electron neutrino with the heaviest eigenmass 
neutrino -- only about $15\%$ above the 
experimental limit (completely allowed order of magnitudewise). In 
this way we predicted successfully $18$ genuinely measured 
parameters not given by the Standard Model itself, and that 
we did with only $6$ adjustable parameters -- the 
vacuum expectation values of Higgs
fields -- in our model. This means that we made $18-6=12$ genuine 
predictions in addition to avoiding limits giving potential problems for most 
interestingly the CHOOZ angle, the 
proton life time and the neutrinoless double $\beta$-decay, however, 
we also did not predict the rate of $\mu\to e +\gamma$ nor similar suppressed 
processes with any problematic rate. If one takes it as we 
mentioned in the earlier article~\cite{FNT} that the smallness of the fine 
structure constants in a model with our gauge group 
is understandable as being really of order 
unity except for the $4\pi$ in $\alpha_i=g_i^2/4\pi$ and the 
corrections from the break down to diagonal subgroup, we can 
claim that we have understood with our model all {\em the 
orders of magnitudes} of the couplings and parameters of the 
Standard Model in terms of only $7$ vacuum expectation values~\cite{MPP}
(if we should claim the VEV of Weinberg-Salam Higgs as 
understood we would, of course, have to include it simply as 
a seventh parameters because we have not explained the 
hierarchy of the weak to Planck scale). 

We did obtain this impressive fit order of magnitudewise 
in a model with the gauge group eq.~(\ref{eq:agut}) and by a 
very data inspired choice of the gauge quantum numbers of our seven 
Higgs fields. We have, however, good reasons to claim that the 
precise detailed structure of our gauge group is not important,
especially in as far as we in reality only needed to use the 
abelian part of the extension of the Standard Model group. The presumably 
only important thing is that the gauge group is sufficiently large, so 
that it separates the various Weyl components of the quarks and the 
leptons, in order to accomplish the various mass matrix element
suppressions needed. On the other hand, the gauge group should 
not be so large as to include $SU(5)$ or other Grand unification 
gauge group, since that tends to give {\em exact} mass relations 
which become a severe problem to avoid, except for the 
$\tau-b$ mass relation. Our gauge group were the biggest 
one represented on the supposed three 
heavy neutrinos and the $45$ already observed Weyl quark and lepton 
components and {\em avoiding such unwanted unifications}. However, 
such a large gauge group is far from being needed although it 
must be considered the suggestion of the present article that a 
relatively large group mainly abelian or, at least, not too much 
``unifying'' is what could fit very well the ``small
hierarchies'' of quark and lepton masses and mixings.

The choices of the specific quantum numbers for our seven 
Higgs fields represents an opportunity of a 
sort of discrete fitting in the sense that we adjust 
these quantum numbers to be able to fit the masses and 
mixings. Historically we developed the quantum number 
proposals making small changes gradually along so as to 
incorporate new experimental data. For example the 
fields $\chi$ and $\phi_{B-L}$ were introduced to cope with the neutrino 
oscillations, and previous Higgs fields (called $S$ and $\xi$) were 
replaced by the fields $\rho$ and $\omega$ in order to fit the Large
Mixing Angel MSW solar neutrino solution, which became favoured 
strongly during the development period of our model.

{}Finally the new step of development in the present article is that 
the quantum numbers of the field $\phi_{B-L}$ got modified to 
make the two heaviest see-saw neutrinos become 
degenerate in mass only deviating by a relatively small 
mass difference 
of the order of the suppression factor associated with the VEV 
of the field $\chi$ being small. Thereby we could enhance the $CP$ 
violation in the decay of the heavier see-saw neutrinos, thus 
achieving a larger baryon number. The latter were needed in as 
far as the previous version of the model, which did not have 
such a degeneracy, gave too little baryon number excess.

Although in the present model we get somewhat more 
wash-out effect we still get 
closer to the from Big Bang fitting estimated baryon number 
density relative to entropy density, and we predict it only a 
factor three (see eq.~\ref{eq:YB}) below the experimental 
data, and that should be counted as only 
$\ln(8.5/2.59)/(\sqrt{10}\cdot0.64) = 0.59$ 
``standard deviations''!

\subsection{How does the model function?}
\indent\ A few comments about how the model functions 
may be on its place: The {\em diagonal} elements in the 
four Dirac mass matrices -- the two for quarks, 
the charged leptons, and the ``neutrino Dirac 
mass matrix'' -- are suppressed to just the same degree in all 
these four matrices. This is so because the quantum number exchanges 
needed to uphold these elements are just those of a 
conventional Standard Model Weinberg-Salam Higgs field, however, 
taken as the family quantum number for that family that 
corresponds to the number of the diagonal matrix element 
in question. This gives at first an order of magnitude 
degeneracy of quarks and leptons in the same family 
at the Planck scale of running couplings, a result that is good for 
simulating $SU(5)$ GUT predictions -- {\em only 
order of magnitudewise} -- and for explaining the crude 
degeneracy of up-quark and electron mass (when extrapolated to 
Planck scale). However, the Charm quark and top-quark masses need 
to be explained as shooting up due to some special mechanism, 
and we managed to get them dominated by off-diagonal elements.
Due to that the first family gets its mass matrix elements 
using $\rho$ and
$\omega$ together with the Higgs fields already used to 
give the second family masses one tends to get just 
the same extra factor of the $\rho^3\sim\omega^3$ type and for 
mixing angles/off-diagonal elements $\omega\rho^\dagger$. There 
is as a result in such models easily achievable 
factorisation of the mixing angles for the quarks and 
for the neutrinos, respectively, at the Planck scale:
\begin{eqnarray}
  V_{u b}\!&\approx&\! V_{u s}\, V_{c b}\nn,\\ 
  \theta_{\rm chooz} \!&\approx&\! \theta_{\odot} \,\, \theta_{\rm atm}\nn.
\end{eqnarray}

This going to first family {\em via} the second 
family just getting an extra $\omega^3$ or $\omega\rho^\dagger$ 
roughly is also behind how in our model we get much less 
hierarchy among the left-handed neutrinos than
for the charged quarks and leptons. In fact the lightness 
of the right-handed
neutrinos ``of first generation'' gives a large 
propagator largely being compensated by the corresponding 
smallness of the first family Dirac mass matrix 
elements. So the much less hierarchy for the neutrinos 
have a rather natural explanation. This fact supports the
see-saw mechanism and thereby the Fukugita-Yanagida scheme.

However, we should admit that the rather 
large neutrino mixing angles in our model 
are obtained by making it possible to fit them to be of 
order unity. We use $\eg$ for the achievement of the 
large atmospheric neutrino mixing angle the adjustment 
of the $\chi$ vacuum expectation value so as to be not 
much different from that of $T$. The solar mixing angle 
is made order unity by having $\rho\sim\omega$ with 
about same vacuum expectation values.

Further mass relations at the Planck scale are in our model:
\begin{eqnarray}
  V_{us}=\theta_{c}\!&\approx&\! \left(\theta_{\odot}\right)^{-\frac{1}{3}}\, 
\left(\frac{m_d}{m_s}\right)^{\frac{2}{3}}\nn,\\
m_b^3 \!&\approx& \! m_s \,\, m_c \,\, m_t \nn.
\end{eqnarray}

Concerning the baryon number production it should be said that
due to the much larger $CP$ violation, $\ie$, $\epsilon_2$ and
$\epsilon_3$, in the decay of the two heavier and almost degenerate 
see-saw neutrinos than in the decay of the lightest one, it is 
the heavy neutrinos that deliver the main/dominant contribution 
to the surviving $(B-L)$ excess and thus to the baryon number.

This of course means that it is important for our success that there
were no inflation after the era of the heavy pair of degenerate see-saw
neutrinos to attenuate the $B-L$ produced and add re-heating entropy 
diminishing $Y_{B-L}$ (or $Y_B$). It is also important for such success 
that we had already at this era roughly the number of see-saw 
neutrinos corresponding to thermal equilibrium as we used in our 
calculation. That means that inflation eras should already have been 
recovered from at the time when the temperature reached down 
to $10^{10}~\GeV$
(see eqs.~\ref{eq:Rmajomass1}-\ref{eq:Rmajomass3}).

\subsection{Developments in calculational method}
\indent\ It should be mentioned that we have made 
the present numerical calculations taking into 
account the renormalisation group running effects 
of the Yukawa couplings as well as the effective dimension 
five operator corresponding to two Weinberg-Salam Higgses 
interacting with two left-handed neutrinos giving the 
neutrino masses observed. Compared to our previous work 
we included the running  between the Planck scale and 
the see-saw scale of the Dirac neutrino Yukawa couplings.

\subsection{On the accuracy}
\indent\ At first one might wonder if such runnings are needed 
when we at the end in principle get only order of magnitude 
predictions. However, they can give easily factors of 
three (or more) and we really have managed to get agreeing results 
with accuracy corresponding to that the simple mass 
matrix elements are uncertain only by the 
theoretically expected $\pm64\%$ logarithmically, so indeed it is important to 
include corrections that could be of the order of a factor two.
It is quite remarkable that we are able to work/fit
with this good accuracy in a model that is in principle only order
of magnitude. If it is not just because we had too many 
discrete details, $\ie$, quantum numbers of Higgs field, to 
adjust it should mean that there is indeed in nature 
couplings that are {\it a priori} of order unity in 
a way that in praxis means that one can count on them 
being of that order to the rather high accuracy we achieved.
With this our $\pm64\%$ accuracy in mind we may remark that we 
have in spite
of working only ``with orders of magnitude'' gotten rather 
close to the experimental uncertainties in the quantities 
which we fit: For the very light quarks up, down and strange 
quarks one can get variations form one lattice computation to 
the other one that can ``well'' reach our uncertainty and the 
neutrino parameters are typically also uncertain with such 
uncertainties.

Of course one can still be more ambitious than we were in 
the present article and hope for understanding the well 
measured mass ratios of the charged leptons, but there 
is not at all as much yet to fit as if the quarks and 
neutrons masses and mixings had been equally well 
measured and thus an order of magnitude fitting is at the 
present time using quite a significant bit of the data 
that can possibly help us to dig behind the Standard Model.

At the end let us admit that our ambitious model does not 
have any candidate in it for dark matter nor for the LSND 
neutrino anomaly. That would require totally new elements 
in our model, as for instance a hidden sector. Combining our model with 
Supersymmetry would enforce us to double all the Higgs fields 
except $W$-Higgs field, and the model would loose much predictive power 
drastically.

\section*{Note added}
\noindent
During the completion of this article, the reference~\cite{AKLR} 
appeared which treats also the $\beta$ function for the Dirac neutrino 
Yukawa coupling. Our result (eq.~\ref{Diracyukawa}) does not match with 
their result, with respect to the coefficient of weakhypercharge 
gauge coupling, because of the different notation: we used the GUT
notation for $g_1$ coupling constant.

\section*{Acknowledgements}
We wish to thank T.~Asaka, W.~Buchm{\"u}ller, L.~Covi and P.~Di Bari 
for useful discussions. It is pleasure to thank all members of 
DESY Theory Group for warm welcome and excellent working environment. 
H.B.N. thanks the Alexander von Humboldt-Stiftung for the Forschungspreis. 
Y.T. thanks DESY for financial support.



\begin{thebibliography}{99}
%
\bibitem{FNT}
C.~D.~Froggatt, H.~B.~Nielsen and Y.~Takanishi, hep-ph/0201152; to be 
published in Nucl.\ Phys.\ B.
%
\bibitem{MSW}
L.~Wolfenstein, Phys.\ Rev.\ D {\bf 17} (1978) 2369;~\ibid{20}{1979}{2634};\\
S.~P.~Mikheev and A.~Yu.~Smirnov, Sov.\ J.\ Nucl.\ Phys.\  {\bf 42} (1985) 913;
Nuovo Cim.\ C {\bf 9} (1986) 17.
%
\bibitem{FN}
C.~D.~Froggatt and H.~B.~Nielsen, Nucl.\ Phys.\ B {\bf 147} (1979) 277.
%
\bibitem{NJL}
Y.~Nambu and G.~Jona-Lasinio, Phys.\ Rev.\  {\bf 124} (1961) 246.
%
\bibitem{seesaw}  T.~Yanagida, in Proceedings of the Workshop on Unified
Theories and Baryon Number in the Universe, Tsukuba, Japan (1979), eds.
O.~Sawada and A.~Sugamoto, KEK Report No. 79-18; \\
M.~Gell-Mann, P.~Ramond
and R.~Slansky in Supergravity, Proceedings of the Workshop at
Stony Brook, NY (1979), eds. P.~van Nieuwenhuizen and D.~Freedman
(North-Holland, Amsterdam, 1979).  
%
\bibitem{NT} 
H.~B.~Nielsen and Y.~Takanishi, Nucl.\ Phys.\ B {\bf 588} (2000) 281;
\ibid{604}{2001}{405}; Phys.\ Lett.\ B {\bf 507} (2001) 241.
%
\bibitem{CKM}
N.~Cabibbo, Phys.\ Rev.\ Lett.\  {\bf 10} (1963) 531;\\
M.~Kobayashi and T.~Maskawa, Prog.\ Theor.\ Phys.\  {\bf 49} (1973) 652.
%
\bibitem{MNS}
Z.~Maki, M.~Nakagawa and S.~Sakata, Prog.\ 
Theor.\ Phys.\  {\bf 28} (1962) 870.
%
\bibitem{pierre}
H.~Arason, D.~J.~Casta{\~n}o, B.~Keszthelyi, S.~Mikaelian, 
E.~J.~Piard, P.~Ramond and B.~D.~Wright, Phys.\ Rev.\ 
D {\bf 46} (1992) 3945.
%
\bibitem{5run}
S.~Antusch, M.~Drees, J.~Kersten, M.~Lindner and M.~Ratz, Phys.\ 
Lett.\ B {\bf 519} (2001) 238; \\
P.~H.~Chankowski and P.~Wasowicz, Eur.\ Phys.\ J.\ 
C {\bf 23} (2002) 249.
%
\bibitem{FF}
C.~D.~Froggatt, H.~B.~Nielsen and D.~J.~Smith, hep-ph/0108262.
%
\bibitem{SNO} Q.~R.~Ahmad {\it et al.}, SNO Collaboration, 
Phys.\ Rev.\ Lett.\  {\bf 87} (2001) 071301.
%
\bibitem{chlorine} B.~T.~Cleveland {\it et al.}, 
Astrophys.\ J.\  {\bf 496} (1998) 505.
%
\bibitem{sage}  J.~N.~Abdurashitov {\it et al.}, SAGE Collaboration,
Phys.\ Rev.\ C {\bf 60} (1999) 055801.
%
\bibitem{gallex} W.~Hampel {\it et al.}, GALLEX Collaboration, 
Phys.\ Lett.\ B {\bf 447} (1999) 127.
%
\bibitem{gno} E.~Belloti, talk at XIX International 
Conference on Neutrino Physics and Astrophysics, Sudbury, 
Canada, June 2000.
%
\bibitem{SK8B}
S.~Fukuda {\it et al.}, Super-Kamiokande Collaboration,
Phys.\ Rev.\ Lett.\  {\bf 86} (2001) 5651.
%
\bibitem{Bahcall}
J.~N.~Bahcall, M.~H.~Pinsonneault and S.~Basu, 
Astrophys.\ J.\  {\bf 555} (2001) 990.
%
\bibitem{fogli}
G.~L.~Fogli, E.~Lisi, D.~Montanino and A.~Palazzo, 
Phys.\ Rev.\ D {\bf 64} (2001) 093007.
%
\bibitem{cc1}
J.~N.~Bahcall, M.~C.~Gonzalez-Garcia and C.~Pe{\~n}a-Garay,
JHEP {\bf 0108} (2001) 014.
%
\bibitem{goswami}
A.~Bandyopadhyay, S.~Choubey, S.~Goswami and K.~Kar, 
Phys.\ Lett.\ B {\bf 519} (2001) 83.
%
\bibitem{smirnov}
P.~I.~Krastev and A.~Yu.~Smirnov, hep-ph/0108177.
%
\bibitem{SKDN}
S.~Fukuda {\it et al.}, Super-Kamiokande Collaboration, 
Phys.\ Rev.\ Lett.\  {\bf 86} (2001) 5656.
%
\bibitem{SK}
Y.~Fukuda {\it et al.}, Super-Kamiokande Collaboration, 
Phys.\ Rev.\ Lett.\  {\bf 81} (1998) 1562; \\
S.~Fukuda {\it et al.}, 
Super-Kamiokande Collaboration, 
Phys.\ Rev.\ Lett.\  {\bf 85} (2000) 3999;
T.~Toshito, talk at the 36th Rencontres de Moriond on Electroweak 
Interactions and Unified Theories, Les Arcs, France, 
10-17 Mar 2001; hep-ex/0105023. 
%
\bibitem{CHOOZ}
M.~Apollonio {\it et al.}, CHOOZ Collaboration, 
Phys.\ Lett.\ B {\bf 466} (1999) 415.
%
\bibitem{cc2}
M.~C.~Gonzalez-Garcia and C.~Pe{\~n}a-Garay, Phys.\ Lett.\ 
B {\bf 527} (2002) 199.
%
\bibitem{Valle}
M.~C.~Gonzalez-Garcia, M.~Maltoni, C.~Pe\~{n}a-Garay 
and J.~W.~Valle, Phys.\ Rev.\ {\bf D 63} (2001) 033005.
%
\bibitem{fogli2}
G.~L.~Fogli, E.~Lisi, A.~Marrone, D.~Montanino and A.~Palazzo, hep-ph/0104221.
%
\bibitem{serguey1}
S.~M.~Bilenky, D.~Nicolo and S.~T.~Petcov, hep-ph/0112216.
%
\bibitem{cecilia}
C.~Jarlskog, Phys.\ Rev.\ Lett.\  {\bf 55} (1985) 1039.
%
\bibitem{evidence}
H.~V.~Klapdor-Kleingrothaus, A.~Dietz, H.~L.~Harney and I.~V.~Krivosheina,
Mod.\ Phys.\ Lett.\ A {\bf 16} (2002) 2409.
%
\bibitem{FY}
M.~Fukugita and T.~Yanagida, Phys.\ Lett.\  {\bf B174} (1986) 45.
%
\bibitem{dibari}
P.~Di Bari, in progress.
%
\bibitem{BACMF}
see for example: M.~Hirsch and S.~F.~King, Phys.\ Rev.\ 
D {\bf 64} (2001) 113005;\\
F.~Buccella, D.~Falcone and F.~Tramontano, Phys.\ Lett.\ 
B {\bf 524} (2002) 241;\\
W.~Rodejohann and K.~R.~Balaji,hep-ph/0201052; \\
G.~C.~Branco, R.~Gonzalez Felipe, F.~R.~Joaquim and 
M.~N.~Rebelo, hep-ph/0202030; \\
T.~Fukuyama and N.~Okada, hep-ph/0202214.
%
\bibitem{'tHooft}
G.~'t Hooft, Phys.\ Rev.\ Lett.\  {\bf 37} (1976) 8; Phys.\ Rev.\  
{\bf D14} (1976) 3432 [Erratum-{\it ibid.}\ D {\bf 18} (1976) 2199].
%
\bibitem{sphaleron}
V.~A.~Kuzmin, V.~A.~Rubakov and M.~E.~Shaposhnikov, 
Phys.\ Lett.\  {\bf B155} (1985) 36.
%
\bibitem{sakharov}
A.~D.~Sakharov, JETP \ Lett.\ {\bf 5} (1967) 24.
%
\bibitem{Luty}
M.~A.~Luty, Phys.\ Rev.\  {\bf D45} (1992) 455.
%
\bibitem{BuPlu}
W.~Buchm{\"u}ller and M.~Pl{\"u}macher, Phys.\ Lett.\  {\bf B431} (1998) 354.
%
\bibitem{CRV}
L.~Covi, E.~Roulet and F.~Vissani, Phys.\ Lett.\  {\bf B384} (1996) 169.
%
\bibitem{API}
A.~Pilaftsis, Int.\ J.\ Mod.\ Phys.\ {\bf A14} (1999) 1811.
%
\bibitem{FHY}
M.~Fujii, K.~Hamaguchi and T.~Yanagida, hep-ph/0202210.
%
\bibitem{KT}
E.~W.~Kolb and M.~S.~Turner, {\it The Early Universe}, %
Addison-Wesley, Redwood City, USA, 1990.
%
\bibitem{HT} J.~A.~Harvey and M.~S.~Turner, Phys.\ 
Rev.\ {\bf D42} (1990) 3344.
%
\bibitem{murayama}
H.~Murayama and A.~Pierce, Phys.\ Rev.\ D {\bf 65} (2002) 055009.
%
\bibitem{MPP}
D.~L.~Bennett and H.~B.~Nielsen, Int.\ J.\ Mod.\ Phys.\ A {\bf 9} (1994) 5155;
{\it ibid.}\ A {\bf 14} (1999) 3313.
%
\bibitem{AKLR}
S.~Antusch, J.~Kersten, M.~Lindner and M.~Ratz, hep-ph/0203233.
%
\end{thebibliography}
\end{document}